\documentclass[aps,pre,twocolumn,superscriptaddress,floatfix]{revtex4-2}
\usepackage{CJK}
\usepackage{graphicx,amsmath,amssymb,color,multirow,xcolor,MnSymbol}
\usepackage{hyperref}
\usepackage{soul,todonotes} 

\begin{document}
	
	\begin{CJK}{UTF8}{mj}
		\title{Extinction drives emergent metastability in complex ecosystems}
		
		\author{Jong Il Park (박종일)}
		\affiliation{Department of Mathematical Sciences, University of Bath, Bath, BA2 7AY, United Kingdom}
		\author{Tim Rogers}
		\email[Corresponding author: ]{ma3tcr@bath.ac.uk}
		\affiliation{Department of Mathematical Sciences, University of Bath, Bath, BA2 7AY, United Kingdom}
		\author{Joseph W. Baron}
		\email[Corresponding author: ]{jwb96@bath.ac.uk}
		\affiliation{Department of Mathematical Sciences, University of Bath, Bath, BA2 7AY, United Kingdom}
		\date{\today}
		
		\begin{abstract}
			Extinction is inevitable; every species eventually dies out, impacting the ecosystem it is part of.
			Over the past few decades, extensive research stemming from the stability-diversity debate has addressed how species diversity contributes to the stability of large ecosystems.
			However, conventional stability criteria often rely on deterministic frameworks, overlooking the intrinsic population fluctuations that allow any species to go extinct by chance.
			In this paper, we incorporate demographic stochasticity into large complex ecosystems with a rule-based model.
			We demonstrate that such demographic fluctuations rapidly prune the low-abundance species from the ecological community, thereby securing higher systemic stability.
			By developing a bottom-up theory to characterise the statistics of extinction dynamics, we discover that the fraction of surviving species exhibits an anomalous heavy-tailed decay over time, revealing the emergence of remnant communities with robust metastability.
			Our results highlight that when demographic fluctuations are accounted for, ecosystems self-stabilise by reducing their diversity even when they are predicted to be chaotic in the deterministic limit.
		\end{abstract}
		
		\maketitle

    \begingroup
    \def\addcontentsline#1#2#3{}

		\section{Introduction}
		Stability in natural ecosystems remains one of the central controversies in theoretical ecology.
		In his seminal work, Robert May~\cite{may1972will} showed through a random matrix approach that linear instability generically emerges in large, complex ecosystems.
		Subsequent theoretical studies confirmed that such instability can give rise to chaotic population dynamics in nonlinear systems~\cite{bunin2017ecological,galla2018dynamically}, yet empirical evidence for chaos has remained scarce~\cite{ellner1995chaos,hastings1993chaos,berryman1989ecological,sibly2007stability}.
		This rarity has been attributed to the difficulty of detecting chaos in practice, owing to short temporal record and observational noise~\cite{sugihara1990distinguishing,barahona1996detection,hunt2003false}.
		For these reasons, chaotic behaviour has long been regarded as something of a theoretical mirage.
		Recent studies, however, have challenged this view, arguing that chaos may be far more common than previously thought, as improved data analysis methodologies reveal its signature more clearly~\cite{rogers2022chaos,beninca2008chaos}.
		
		In parallel, theoretical efforts have also been devoted to understanding more clearly how many-species ecological communities can remain in the vicinity of a stable fixed point, rather than exhibiting chaos. 
		In part, this has been accomplished by incorporating the multifaceted nature of real ecosystems in more detail.
		These studies have elucidated various mechanisms through which ecosystem stability is modulated by factors such as ecological network structures~\cite{park2024incorporating,poley2025interaction,aguirre2024heterogeneous,emary2022stability}, spatial heterogeneities~\cite{gravel2016stability,baron2020dispersal}, higher-order interactions~\cite{bairey2016high,grilli2017higher}, and nonlinear feedbacks~\cite{hatton2024diversity,neutel2016linking}.
		While the deterministic models considered in these works offer clear analytical virtues under this restriction, they suffer from an inherent limitation: an inability to incorporate the intrinsic stochasticity of real-world population dynamics. 
		As has been noted in previous works, there exist a variety of mechanisms via which noise can alter equilibrium/stability properties \cite{d2005noise,parker2011noise} or suppress chaos \cite{yamazaki1998can}. 
		
		Although recent studies of many-species systems have sought to integrate stochasticity into their modelling frameworks, they frequently face limitations; either they impose arbitrary abundance thresholds to artificially force extinctions~\cite{altieri2021properties,al2026spatiotemporal}, or they treat noise as an externally imposed parameter rather than an intrinsic feature of the system~\cite{garcia2024interactions,de2025self}. 
		Building upon the approach suggested in Ref.~\cite{larroya2023demographic}, we introduce a rule-based model wherein demographic noise is implicitly embedded within the microscopic rules. 
		This allows us to study extinction events in a less contrived way, and therefore understand how noise-induced extinctions affect stability (and therefore the presence of chaos) in many-species communities.
		
		More specifically, in the spirit of May's random matrix approach, we consider a large ecosystem with randomly interacting species. 
		Through numerical simulations, we first investigate the population dynamics, observing qualitative differences between our stochastic model and its deterministic counterpart (the generalised random Lotka-Volterra equations-- grLV).
		Specifically, we find that a rapid reduction in species diversity driven by the intrinsic stochasticity allows the system to secure its stability, even within parameter regimes that correspond to deterministic chaos. 
		In doing so, we uncover that chaotic ecosystems can stabilise themselves without the aid of any external influence.
		
		In order to deepen our understanding of the underlying mechanisms governing this `self-stabilisation' process, we further conduct a theoretical analysis, combining a system-size expansion \cite{vankampen2007stochastic} with dynamic mean-field theory \cite{mezard1988spin, roy2019numerical, altieri2020dynamical}. 
		That is, we first derive the approximate stochastic differential equations (SDEs) that govern the evolution of the population under our proposed rules. 
		In so doing, we effectively add demographic noise to the grLV in a way that accurately reflects the underlying rule-based model. 
		Since we seek the macroscopic community behaviour regardless of the fine-grained details of specific interactions, we then further employ a mean-field reduction. 
		This allows us to consider a single `test' species interacting with a self-consistent environment formed by the rest of the species, thus greatly simplifying the analysis of the many-species system.
		This puts us in a position to evaluate the statistics of the rare, noise-induced extinction events for each species by identifying the most probable paths to extinction. 
		
		We demonstrate that our approximate theory is in quantitative agreement with the original rule-based model.
		Furthermore, our analytical framework sheds light on the prolonged persistence of the surviving community.
		A key finding is that, after the initial rapid extinctions that extinguish the chaotic behaviour, the subsequent fraction of surviving species decays as a power law over time. 
		This indicates a far more robust sustainability than the exponential decay typically observed in few-species rule-based models.
		By analytically evaluating the decay exponent, we show that this power-law decay arises from a non-trivial interplay between random interactions and intrinsic noise, and we confirm that a catastrophic collapse does not occur within computationally observable timescales in our many-species model.
		These findings suggest that intrinsic demographic noise may in part play a constructive role in sustaining stability in natural ecosystems. 
		
		\section{Results}
		
		\subsection{Extinction drives self-stabilisation}
		\begin{figure}[t!]
			\centering
			\includegraphics[width=\linewidth]{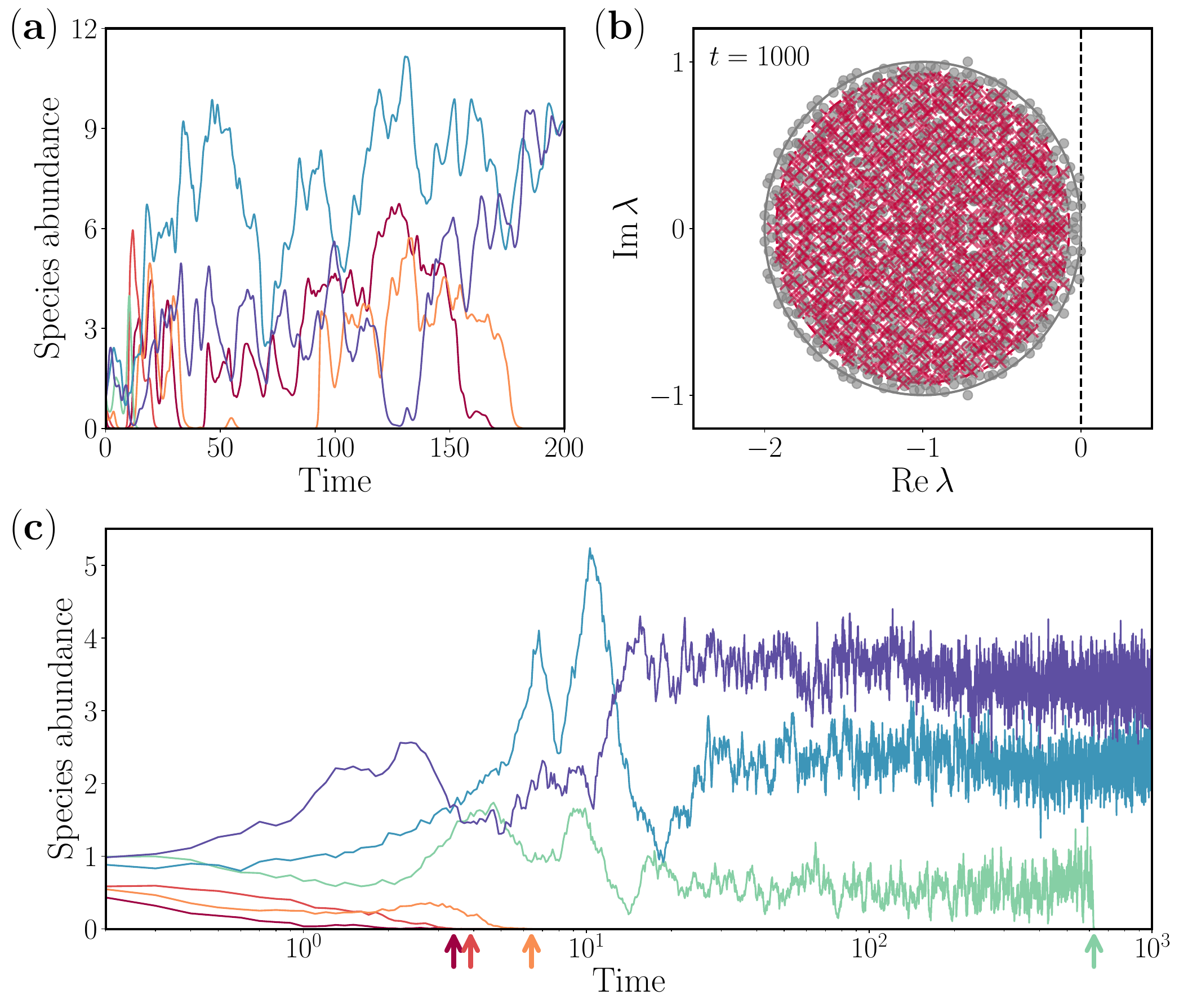}
			\caption{\textbf{Distinct dynamical behaviours and eigenvalue spectra of the surviving species.} \textbf{(a)} Deterministic $(V\rightarrow\infty)$ grLV [see Eq.~(\ref{Seq:deterministic})] \textbf{(c)} rule-based dynamics [see Eq.~(\ref{eq:rules})] for six representative species. Both models are simulated with an identical interaction matrix $\mathbf{J}$, with consistent colour-coding across panels for each species. \textbf{(b)} Eigenvalue spectrum of the reduced interaction matrices $\mathbf{J}^\star$ (see text for definition) for the surviving communities at time $t=1000$. System parameters are chosen such that May's stability criterion predicts chaotic behaviour. While the spectrum of the deterministic model (grey circles) reaches the stability boundary (zero), the rule-based model (red crosses) exhibits the contracted spectrum due to additional extinctions triggered by demographic fluctuations. Notably, while the deterministic system is expected to possess exponentially many unstable fixed points  \cite{ros2023generalized}, only marginally stable states are numerically observable. Parameters: $S = 4096$, $V = 2000$, $\mu = -3$, $\sigma = 2$.}
			\label{fig:fig1}
		\end{figure}
		
		Rule-based modelling is a canonical approach for incorporating demographic fluctuations into population dynamics. The rule-based model (RBM) that we focus on in this work (see Methods Section \ref{sec:RBM} for a full description) is chosen such that we recover the classic many-species Lotka-Volterra equations \cite{bunin2017ecological, galla2018dynamically} when each species is very abundant, and the noise is therefore small. 
		The rule-based model can therefore be seen as incorporating stochasticity into the many-species Lotka-Volterra equations in a bottom-up fashion. 
		
		To be more precise, let us consider a middle ground between the RBM and the deterministic Lotka-Volterra dynamics, which is obtained when the abundances of each species are large but finite. Under these circumstances, each species abundance $x_i(t)$ (where $i = 1, \dots, S$) can be approximated as obeying the following Langevin dynamics \cite{krumbeck2021fluctuation,larroya2023demographic}
		\begin{align}
			\dot{x}_i = x_i \left[1 - x_i + \sum_{j = 1}^S J_{ij} x_j \right] + \sqrt{\frac{B_i(\mathbf{x};\mathbf{J})}{V}}\xi_i(t),
			\label{eq:Kurtz}
		\end{align}
		where $\xi_i(t)$ is a standard Gaussian white noise satisfying $\langle \xi_i(t)\xi_j(t')\rangle = \delta_{ij}\delta(t-t')$ and $B_i(\mathbf{x};\mathbf{J})$ is the state-dependent noise amplitude derived from the reaction rules (see Eq.~(\ref{Seq:Kurtz}) in Methods).
		We consider a Gaussian random matrix for $\mathbf{J}$ whose mean and variance are given by $\langle J_{ij}\rangle_J = \mu/S$ and $\langle J_{ij}^2 \rangle_J - \langle J_{ij}^2 \rangle_J = \sigma^2/S$.
		One thus sees that the usual many-species Lotka-Volterra equations are recovered for $V \to \infty$, where $V$ dictates the typical size of each species' population. 
		
		We emphasise that while Eq.~(\ref{eq:Kurtz}) will form the backbone of our analysis in this paper, the results of simulations in the figures, to which we compare our theory, are always generated by the RBM (unless stated otherwise). 
		Let us now compare the results of the RBM with those of the corresponding deterministic model, which is obtained by setting $V \to \infty$ in Eq.~(\ref{eq:Kurtz}).
		
		\begin{figure*}[t!]
			\centering
			\includegraphics[width=\linewidth]{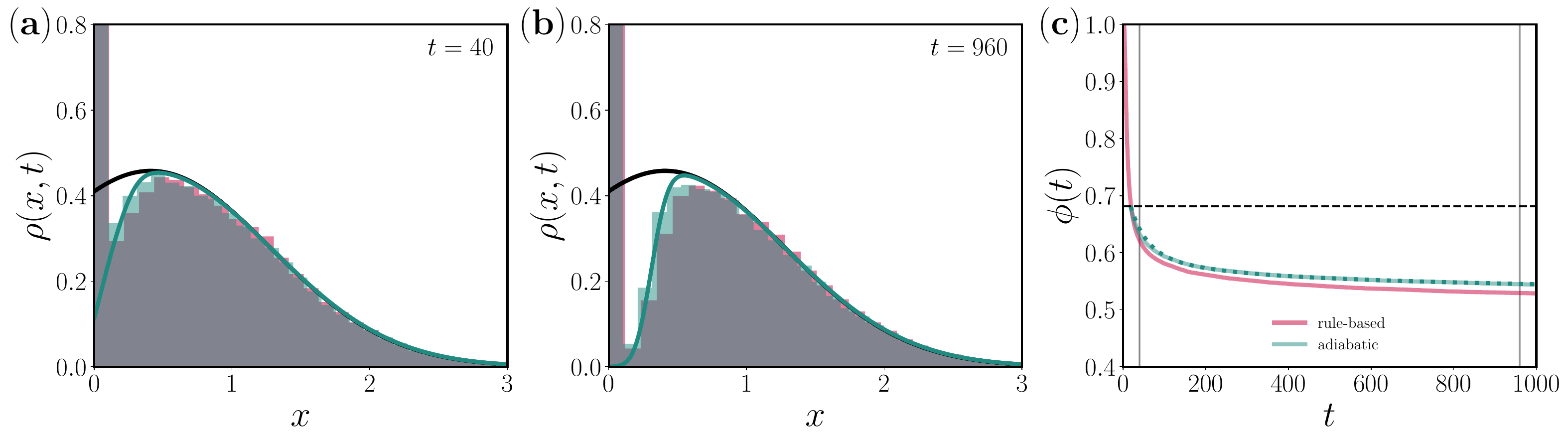}
			\caption{\textbf{Abundance distributions and survival probabilities over time.} Abundance distributions $\rho(x,t)$ obtained from rule-based (red) and adiabatic (green) dynamics at \textbf{(a)} $t = 40$ and \textbf{(b)} $t=960$. Histograms are computed from the numerical simulations of the rule-based [see Eq.~(\ref{eq:rules})] and adiabatic [see Eq.~(\ref{Seq:FP})] dynamics. Solid lines represent theoretical predictions from the deterministic fixed point (black) [see Eq.~(\ref{Seq:ssad})] and adiabatic approximation (green) [see Eq.~(\ref{eq:qsad})]. \textbf{(c)} Survival probabilities $\phi(t)$ as a function of time. Shaded areas indicate the time windows corresponding to the distributions in \textbf{(a)} and \textbf{(b)}. Dashed and dotted lines are predictions from deterministic and adiabatic theories, respectively. The adiabatic results are shifted by $t_d = 18.5$, a value estimated from the rule-based simulation. All results are averaged over 30 independent configurations. Parameters: $S = 1024$, $V=1000$, $\mu = -1$, $\sigma = 1$.}
			\label{fig:fig2}
		\end{figure*}
		
		By accounting for demographic noise, a previous study \cite{larroya2023demographic} reported that ecological communities become more vulnerable to fluctuations when the heterogeneity of interactions is high.
		On the contrary, however, we observe that such extinction risk can ultimately enhance the overall stability of the ecosystem.
		In the regime predicted to be chaotic in the deterministic limit, we compare the dynamical behaviours of the two models (Fig.~\ref{fig:fig1}).
		While the deterministic model displays complex, aperiodic dynamics as we expected from deterministic theory, the rule-based system enters a stabilised phase where species abundances fluctuate around persistent values (blue-toned lines in Fig.~\ref{fig:fig1}a, c).
		
		This self-stabilisation can be understood via May's stability criterion~\cite{may1972will,allesina2012stability}.
		In the limit $S\rightarrow \infty$, the classic theory states that the eigenvalue spectrum of the reduced interaction matrix $\mathbf{J}^\star$ for the surviving species (made by removing the rows and columns from $\mathbf{J}$ that correspond to extinct species) forms a disk in the complex plane with a radius of $\sigma\sqrt{\phi}$, where the fraction of surviving species $\phi$.
		Linear stability is determined by the maximal eigenvalue of $\mathbf{J}^\star$ \cite{baron2023breakdown}, leading to an instability transition at $\sigma^2\phi = 1$.
		Thus, for our rule-based model, demographic noise drives low-abundance species to vanish at the initial stage of dynamics (red-toned lines in Fig.~\ref{fig:fig1}c), thereby securing linear stability by reducing the $\phi$ and therefore the maximal eigenvalue.
		We verify this by measuring eigenvalues at time $t=1000$: the deterministic model exhibits marginal stability with the eigenvalue spectrum just grazing the imaginary axis (in a similar way to what was pointed out in Ref. \cite{biroli2018marginally}), whereas the rule-based model manifests a more contracted spectrum, indicating the emergent stability (Fig.~\ref{fig:fig1}b).
		
		\begin{figure*}[t!]
			\centering
			\includegraphics[width=\linewidth]{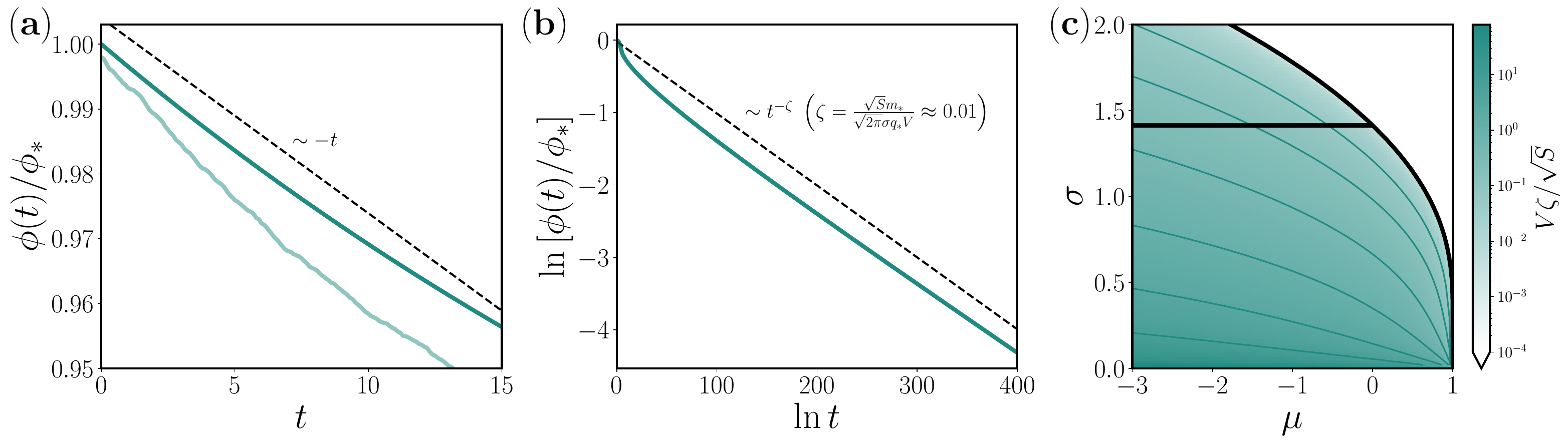}
			\caption{\textbf{Asymptotic behaviours of extinction dynamics and decay exponents.} Survival probabilities $\phi(t)$ at \textbf{(a)} short and \textbf{(b)} long timescales. The green transparent line represents results from the adiabatic simulation [see Eq.~(\ref{Seq:FP})], while the green solid lines indicate numerical solutions of the theoretical prediction [see Eq.~(\ref{eq:qsad})]. Black dashed lines denote the predicted asymptotic limits. \textbf{(c)} Decay exponents scaled by $\sqrt{S}/V$ for the long-time asymptotes in the $(\mu,\sigma)$-plane. The green solid lines indicate contours of $\ln(V\zeta/\sqrt{S})$, and the black solid lines indicate the phase boundaries predicted for deterministic grLV systems. Parameters for \textbf{(a)} and \textbf{(b)}: $S=1024$, $V=1000$, $\mu=-1$, $\sigma=1$.}
			\label{fig:fig3}
		\end{figure*}
		
		It is important to note that the nature of this self-stabilisation process differs fundamentally from deterministic stability.
		Even when the system appears stabilised, demographic noise ensures that extinction occurs rarely but inevitably over long timescales (green line in Fig.~\ref{fig:fig1}c).
		Nevertheless, we show through our theoretical framework that such extinction dynamics are extremely slow, validating the existence of a persistent surviving community.
		This robust persistence is a definitive signature of emergent `metastability'. 
		
		\subsection{Route to the effective dynamics}
		Building on the preceding phenomenological investigation, we now provide a theoretical analysis to deepen our quantitative understanding.
		Beginning with Eq.~(\ref{eq:Kurtz}), our objective is to find a single effective stochastic process that describes the statistics of each species in the many-species system. 
		This is accomplished by carefully handling the random interaction coefficients $J_{ij}$, with the application of a so-called dynamical mean-field theory approach (DMFT) \cite{altieri2020dynamical, galla2018dynamically}.
		
		Unlike the standard grLV framework, the noise amplitude $B_i$ in our model is dependent on the absolute interaction strengths $|J_{ij}|$, requiring a more careful DMFT treatment.
		This interdependency introduces non-Gaussian features in the effective dynamics. However, these are negligible by considering only leading-order terms in $S$ and $V$.
		Furthermore, given the scaling $B_i\sim O(\sqrt{S})$ [see Eq.~(\ref{eq:Kurtz})], we  must consider the atypical thermodynamic limit $S,V \rightarrow \infty$, but with $\sqrt{S}/V$ small and finite. 
		We emphasise that this parameter regime is commonly observed in real ecosystem.
		Since $V$ is analogous to biomass or population per species, the ratio $\sqrt{S}/V$ has been empirically measured to be small across diverse ecological communities, typically ranging from $10^{-4}$ to $10^{-1}$~\cite{white2012characterizing}.
		Additionally, scaling relations reported for microbial communities~\cite{locey2016scaling} show that this ratio scales as $\sim S^{-1.35}$, reinforcing the notion that complex ecosystems generally operate in a small-noise regime.
		
		We thus finally obtain the following effective dynamics
		\begin{align}
			\dot{x} = x\left[1-x+\mu m(t) + \sigma \eta(t)\right]+\sqrt{\left(\sqrt{\frac{2S}{\pi} }  \frac{\sigma m }{V}x\right)} \,\xi(t).
			\label{eq:dmft}
		\end{align}
		We thus see that the interaction term $\sum_{j=1}^SJ_{ij} x_j$ in Eq.~(\ref{eq:Kurtz}) has been replaced with an average interaction $\mu m(t)$ with the mean abundance $m(t)\equiv \langle x(t)\rangle$ and a centred Gaussian coloured noise $\eta(t)$, whose correlator is self-consistently determined as $\langle \eta(t)\eta(t')\rangle = \langle x(t)x(t')\rangle$, which reflects the heterogeneity in species interactions. 
		Here, the angular bracket without subscript $\langle \cdots \rangle$ indicates an averages over both the interaction matrix entries $J_{ij}$ and the demographic noise. 
		We note that the variance of the Gaussian white noise $\xi(t)$, which instead comes from the intrinsic noise from the rule-based model, also depends on $m(t)$, and $\sqrt{S}/V$ serves as the sole parameter that explicitly controls its magnitude (see Methods for full derivation). 
		
		\subsection{Power-law decay of the surviving fraction of species}
		Even with the simplification made by the DMFT, the entanglement of the coloured noise $\eta(t)$ and the white noise $\xi(t)$ makes direct analytical solution elusive.
		To overcome this difficulty, we focus on the small-demographic-noise regime ($\sqrt{S}/V\ll 1)$, where physical intuition suggests that the system should remain in the vicinity of the deterministic stable fixed point for an exponentially long time \cite{hanggi1990reaction}. 
		This persistence allows us to distinguish the extinction dynamics occurring before and after the system reaches its fixed point. 
		We therefore employ an `adiabatic' ansatz that exploits the separation of timescales between the self-consistent noise $\eta(t)$ and demographic fluctuations $\xi(t)$.
		
		More precisely, we assume that the dynamics in Eq.~(\ref{eq:dmft}) rapidly converges to the deterministic fixed point during the initial stage.
		At the end of this stage, the stationary abundance distribution follows the well-known truncated Gaussian~\cite{bunin2017ecological,galla2018dynamically} $\rho_*(x) = \frac{1}{\sqrt{2 \pi \sigma^2 q_*}} \exp\left\{-\frac{(x-1-\mu m_*)^2}{2\sigma^2 q_*} \right\}$ for $x>0$. The static order parameters can be evaluated self-consistently from this expression: $m_* = \int_{0^+}^\infty dx\, x \rho_*(x)$ and $q_*  = \int_{0^+}^\infty dx\, x^2 \rho_*(x)$.
		
		Once the system reaches this fixed point are driven primarily by demographic fluctuations.
		In the small-noise regime, these processes are expected to be significantly slower than the initial extinctions driven by competitive interactions during the previous relaxation phase. 
		Such rare extinctions can effectively be treated by applying the Wentzel-Kramers-Brillouin (WKB) ansatz~\cite{assaf2017wkb,bressloff2022wkb}, yielding the post-stationary abundance distribution over time (see Methods for full derivation): 
		\begin{equation}
			\begin{aligned}
				\rho(x,t) &\approx [1-\phi(t)]\delta(x) + \Theta(x)e^{-A(x,t)},\\
				A(x,t) &= \frac{t}{\tau_e(x)} + \frac{[x-1-\mu m(t)]^2}{2\sigma^2 q(t)} + \frac{1}{2} \ln{[2\pi\sigma^2 q(t)]},
				\label{eq:qsad}
			\end{aligned}
		\end{equation}
		where $\Theta(x)$ is the Heaviside step function and the survival probability $\phi(t) \equiv \langle \Theta(x)\rangle=\int_{0^+}^\infty dx\,\rho(x,t)$ denotes the fraction of surviving species at time $t$.
		The quasi-static order parameters are again self-consistently defined, $m(t) = \int_{0^+}^\infty dx\,x\rho(x,t)$ and $q(t) = \int_{0^+}^\infty dx\,x^2\rho(x,t)$ with initial conditions $m(0) = x_*$ and $q(0) = q_*$.
		In Eq.~(\ref{eq:qsad}), the mean extinction time  is characterised as a super-exponentially increasing function of $x$, given by
		\begin{align}
			\tau_e(x) = \frac{\pi}{x}\mathrm{erfi}\left(\frac{\sqrt{V}x}{\sqrt{(2S/\pi)^{1/2}\sigma m(t)}}\right)
			\label{eq:met}
		\end{align}
		where the imaginary error function has the asymptotic expansion $\mathrm{erfi}(x)\sim e^{x^2}/(\sqrt{\pi}x)$ as $x \rightarrow \infty$.
		It is straightforward to verify that the usual deterministic solution \cite{bunin2017ecological,galla2018dynamically} is recovered at $t=0$, \textit{i.e.} $\rho(x,0) = \rho_*(x)$.
		
		Motivated by our phenomenological observations in Fig.~\ref{fig:fig1} and the analytical findings in Eq.~(\ref{eq:met}), we further neglect the temporal variation of the order parameters by approximating $m(t) \approx m_*$ and $q(t)\approx q_*$.
		This simplification remains valid over extended timescales because post-stationary extinctions predominantly involve low-abundance species, thus exerting a minimal influence on the collective statistics of the system. 
		
		We test the adiabatic approximation for $\rho(x,t)$ and $\phi(t)$ against numerical simulations of the rule-based model in Fig.~\ref{fig:fig2}, where we observe a far better agreement compared to the deterministic theory, particularly for low-abundance species.
		Crucially, the validity of these results relies on the assumption that the perturbations to the order parameters induced by the gradual species loss remain small.
		
		Drawing upon our adiabatic theory, we further investigate the characteristics of the post-stationary extinction dynamics.
		From the analytical form of the mean extinction time in Eq.~(\ref{eq:met}), the extinctions of low-abundance species dominate at early times $(t\ll 1)$.
		In this regime, the mean extinction time remains nearly independent of species abundance, resulting in a linear decay of the survival probability $\phi(t) \simeq \phi_*(1-\alpha t)$, where $\alpha$ is a small parameter determined by the deterministic order parameters.
		
		Following the depletion of low-abundance species, the remaining populations (when stratified by time averaged abundance $x$) each possess exponentially large mean extinction times, signifying the emergence of a quasistationary state.
		Remarkably, in the long-time limit $(t \gg 1)$, this extinction process of the full community (integrating over abundance $x$) is significantly slower than standard Kramers' escape rate, exhibiting a power-law decay in the survival probability
		\begin{align}
			\phi(t)\propto t^{-\zeta}\quad\text{with}\quad \zeta = \frac{\sqrt{S}m_*}{\sqrt{2\pi}\sigma q_*V}.
			\label{eq:decay}
		\end{align}
		The decay exponent $\zeta$ serves as a representative macroscopic observable of our model.
		From a physical perspective, the quasistationary abundance distribution can be expressed as the superposition of two Gaussian kernels: one representing the random interactions (with variance $\sigma^2 q_*$) and the other stemming from the demographic fluctuations (with variance $\sqrt{S}\sigma m_*/(\sqrt{2\pi}V)$).
		Interestingly, the exponent $\zeta$ is determined by the ratio of these two variances, demonstrating that the power-law decay of  $\phi(t)$ originates from the interplay between quenched disorder and demographic noise.
		While there are subleading contributions to $\zeta$ that depend on time $t$, these vanish in the infinite-time limit, \textit{i.e.} $\lim_{t\rightarrow \infty} \zeta(t) = \zeta$ (see Methods for details). 
		Perhaps there may be some parameter regimes where the adiabatic approximation breaks down and non-trivial feedback affects the dynamics more significantly. 
		While intriguing, accounting for this coupling is beyond the scope of the present work.
		
		We compare these asymptotic formulae with the numerical evaluation of $\phi(t)/\phi_*$ obtained via direct integration of $\rho(x,t)$.
		By examining the scaled exponent $V\zeta/\sqrt{S}$ across the $(\mu,\sigma)$ parameter space, we find that it vanishes $(\zeta \rightarrow 0)$ at the onset of the unbounded growth phase, where the divergence of $q_*/m_*$ leads to a dynamical freezing of the extinction process (Fig.~\ref{fig:fig3}).
		
		\section{Discussion}
		In this study, we have unveiled the impact of intrinsic demographic noise on ecosystem stability.
		By establishing a rule-based model, stochasticity is incorporated naturally into the underlying population dynamics, revealing that the noise-driven extinction of low-abundance species actually promotes ecosystem stability from the perspective of May's framework.
		Furthermore, we have developed and applied the adiabatic approximation during our analytical treatment. 
		This not only provided a more accurate prediction for the abundance distributions compared to the deterministic theory, but crucially it also allowed us to show that the fraction of surviving species exhibits power-law decay over time, concretising the mechanism of the self-stabilisation process.
		
		We emphasise the ecological significance of our work in addressing a long-standing puzzle: why chaos is rarely observed in natural ecosystems~\cite{sibly2007stability,ellner1995chaos}.
		We propose that large ecosystems can be self-stabilising, via the extinction of low-abundance species, rendering chaos a purely transient phenomenon when this occurs.
		Furthermore, we found that these reduced communities are surprisingly robust over time, despite the presence of noise and the absence of stabilising features such as species dispersal in space~\cite{gravel2016stability} or predator-prey interactions~\cite{allesina2012stability, galla2018dynamically}.
		Since our analysis remains consistent with May's stability criterion, we believe our observations go some way towards reconciling May's predictions with ecological reality.
		
		Our study also opens the door to understanding a range of other noise-induced phenomena in complex ecosystems and disordered systems more generally. 
		In particular, while we have considered the mean-field limit in which species are essentially treated as independent, it would be fascinating to understand the rare collective fluctuations that could lead to the extinction events of many species simultaneously. 
		Such an investigation could be facilitated by studying the instantons (the most likely paths) of the system again via path-integral approaches similar to ours. 
		Indeed, such instanton analyses have had success in the area of spin-glass physics for understanding how physical systems navigate complex energy landscapes~\cite{ros2021dynamical,lopatin1999instantons}.
		In the ecological context, this would allow one not only to understand the extinction dynamics of individual species, but also the routes by which collections of species undergo catastrophic collapse, and what kinds of communities are most likely to exhibit such fluctuations.
		
		Another way to gain insight into species' vulnerability to extinction is to extend our simple model in such a way that species become more statistically distinct. 
		That is, in the present model, the species interactions are fully random, and so every species is in some sense like any other. 
		If we were to introduce further model aspects such as an interaction network (i.e. each species only interacts with a select few others)~\cite{park2024incorporating, aguirre2024heterogeneous, poley2025interaction}, a species hierarchy or trophic levels (such as in the niche or cascade models)~\cite{poley2023generalized, grilli2016modularity, allesina2015predicting}, or differential birth/death/self-regulation rates \cite{barabas2017self}, we could then understand in more depth what characteristics (e.g.~network centrality/connectivity, hierarchy position, or birth/death rate) make a particular species most vulnerable to extinction. 
		
		As a further extension of our simple model, one could also consider including other interaction types, such as cooperative, competitive or predator-prey interactions, rather than the simpler commensalistic/ammensalistic interactions that were considered here (see Methods). 
		Predator-prey interactions such as $X_i + X_j \rightarrow 2X_i$~\cite{krumbeck2021fluctuation} for example, naturally gives rise to demographic noise correlations between species.
		It would be intriguing to investigate how such correlations impact the metastability of the ecosystem.
		
		As a final remark, the theoretical framework developed here is not restricted to ecological systems, but extends to generic complex birth-death processes.
		Previous studies have largely focused on one of two extremes: stochastic population dynamics with few species, or complex interactions in the deterministic limit.
		Our work bridges the gap between these two regimes, suggesting that the present approach may serve as a versatile tool for analysing rare-event dynamics in large, complex stochastic systems.
		
		\section{Methods}
		\subsection{Rule-based model}\label{sec:RBM}
		We consider a system of many species that we label $i = 1,\cdots,S$. 
		Given some initial abundances of each species, each of the following events may occur probabilistically
		\begin{equation}
			\begin{aligned}
				&X_i \xrightarrow{\lambda_b}2X_i,\quad X_i \xrightarrow{\lambda_d} \emptyset&\text{(intrinsic birth/death)},\\
				&2X_i \xrightarrow{1/V} X_i&\text{(self-regulation)},\\
				&X_i+X_j \xrightarrow{J_{ij}^-/V}X_j&\text{(amensalistic interaction)},\\
				&X_i+X_j\xrightarrow{J_{ij}^+/V} 2X_i+X_j&\text{(commensalistic interaction)}.
			\end{aligned}
			\label{eq:rules}
		\end{equation}
		Here, the notation $A + B \xrightarrow{\lambda} C+D$ indicates an event where individuals of species $A$ and $B$ interact to produce $C$ and $D$ with a probability per unit time $\lambda$. 
		We thus see that each species undergoes birth and death events independently of the other species. 
		The inter-species interactions are encapsulated by random interaction coefficients $J_{ij}$, which are drawn independently from a Gaussian distribution with mean $\langle J_{ij}\rangle_J = \mu/S$ and variance $\langle J_{ij}^2 \rangle_J - \langle J_{ij} \rangle_J^2 = \sigma^2/S$. Here, $\langle \cdots \rangle_J$ indicates an average over the fixed random matrix entries (to be distinguished from averages over other sources of randomness, such as demographic noise). 
		Defining $J_{ij}^+ = \max(J_{ij},0)$ and $J_{ij}^- = \max(-J_{ij},0)$, we imagine that species pairs $(i,j)$ with positive $(J_{ij}, J_{ji})$ have commensalistic interactions (i.e. species $i$ benefits, and nothing happens to species $j$), while those with negative coefficients have amensalistic interactions (species $i$ suffers, and nothing happens to species $j$). 
		In principle, we could include other interactions types, such as cooperation, competition or parasitism, but this construction is the simplest for our analysis. 
		We note that the parameter $V$ (known as the system size) controls the relative magnitude of the noise in the system, with population counts scaling as $V$ and fluctuations scaling as $\sqrt{V}$. 
		Under this construction, the species abundance (the population scaled by typical size $V$) evolves according to the usual generalised random Lotka-Volterra (grLV) equation~\cite{bunin2017ecological,galla2018dynamically} [see also Eq.~(\ref{Seq:deterministic})] in the deterministic limit $V\rightarrow \infty$, as we show in the next subsection.
		
		\subsection{Effective dynamics}
		\label{sec:eff}
		In this section, we present the derivation of the effective dynamics.
		Starting from the microscopic reaction rules, we formulate  equivalent SDEs in the continuum limit and subsequently derive the corresponding effective single-species dynamics via DMFT. A full account of this calculation is provided in the Supplemental Material. Here, we discuss the main steps. 
		
		For the stochastic dynamics of the populations $n_i$ (i.e. the raw numbers of each species $i$) obeying the rules in Eq.~(\ref{eq:rules}), the joint probability distribution $P(\mathbf{n},t)$ is governed by following chemical master equation~\cite{vankampen2007stochastic,gardiner2009stochastic}
		\begin{equation}
			\begin{aligned}
				&\partial_t P(\mathbf{n},t) = \\
				&\qquad\sum_i \left[(\mathcal{E}_i^- - 1)W^+(\mathbf{n})  +(\mathcal{E}_i^{+}-1)W^-(\mathbf{n}) \right]P(\mathbf{n},t)
				\label{Seq:CME}
			\end{aligned}    
		\end{equation}
		with transition rates
		\begin{equation}
			\begin{aligned}
				W^+(\mathbf{n}) &= \lambda_b n_i + \sum_{j\in \partial i^+}J_{ij}^+ \frac{n_i n_j}{V},\\
				W^-(\mathbf{n}) &= \lambda_d n_i + \frac{n_i(n_i -1)}{V} + \sum_{j\in \partial i^-}J_{ij}^- \frac{n_i n_j}{V},
			\end{aligned}    
		\end{equation}
		where $\mathcal{E}_i^{\pm}:n_i\mapsto n_i\pm 1$ denote the step operators, and $\partial i^{\pm}$ represent disjoint sets of neighbours of $i$ connected by the assigned interactions $J^{\pm}_{ij}$.
		
		In the continuum limit $(1\ll V<\infty)$, we approximate Eq.~(\ref{Seq:CME}) as a set of Langevin equations by introducing the continuous abundance $x_i \equiv n_i/V$.
		In accordance with Kurtz's theorem~\cite{kurtz1978strong}, the approximated process is given by
		\begin{equation}
			\begin{aligned}
				\dot{x}_i &= x_i \left(\lambda_b - \lambda_d - x_i + \sum_{j} J_{ij} x_j \right) \\
				&\qquad+ \sqrt{V^{-1} x_i\left(\lambda_b + \lambda_d +x_i+\sum_{j}|J_{ij}|x_j\right)}\xi_i(t),
				\label{Seq:Kurtz}
			\end{aligned}    
		\end{equation}
		where $\xi_i(t)$ is a centred, unit-variance Gaussian white noise.
		This equation is a full expression of Eq.~(\ref{eq:Kurtz}) in the main text. We also preset a derivation of Eq.~(\ref{Seq:Kurtz}) using a system-size expansion in the Supplemental Material. 
        
		Without loss of generality, we hereafter set $\lambda_b = 1$ and $\lambda_d \approx 0$. 
		One notes that in the deterministic limit $V \to \infty$ we recover the usual many-species Lotka-Volterra equations
		\begin{align}
			\dot x_i = x_i \left( 1 - x_i + \sum_j J_{ij} x_j\right). 
			\label{Seq:deterministic}
		\end{align}
		
		While more manageable than the chemical master equation, the description in Eq.~(\ref{Seq:Kurtz}) still necessitates solving $S$ coupled SDEs.
		Given that our interest lies in the collective statistics of the system, we focus on the representative dynamical behaviour of the species rather than individual trajectories.
		By assuming statistical equivalence, whereby species are treated as indistinguishable, the high-dimensional dynamics can be reduced to a single effective SDE.
		Following this conceptual outline of the DMFT, we employ the Martin-Siggia-Rose-de Dominicis-Janssen (MSRDJ) field-theoretic formalism to carry out the procedure.
		We first construct the MSRDJ moment-generating functional as
		\begin{align}
			Z[\boldsymbol{\psi}](\mathbf{J}) \equiv \int \mathcal{D}[\mathbf{x},\hat{\mathbf{x}}]\exp \{A_0[\mathbf{x},\hat{\mathbf{x}}] + A_1[\mathbf{x},\hat{\mathbf{x}}](\mathbf{J})\}
		\end{align}
		with actions
		\begin{equation}
			\begin{aligned}
				A_0[\mathbf{x},\hat{\mathbf{x}}] &= i\sum_i\int dt\,x_i(t)\psi_i(t)\\
				&+ i\sum_i\int dt\,\hat{x}_i(t)\left\{\dot{x}_i(t) -x_i(t) \left[1 - x_i(t) \right]\right\}\\
				&-\frac{1}{2V}\sum_i\int dt\,\hat{x}_i(t)^2x_i(t)\left[1+x_i(t)\right]
			\end{aligned}
		\end{equation}
		and
		\begin{equation}
			\begin{aligned}
				A_1[\mathbf{x},\hat{\mathbf{x}}](\mathbf{J}) &=-i\sum_{i,j}\int dt\,\hat{x}_i(t)J_{ij}x_i(t)x_j(t)\\
				&\quad-\frac{1}{2V}\sum_{i,j}\int dt\,\hat{x}_i(t)^2|J_{ij}|x_i(t)x_j(t),
			\end{aligned}
		\end{equation}
		where $\int \mathcal{D}[\mathbf{x},\hat{\mathbf{x}}]$ denotes an integral over all paths $(\mathbf{x},\hat{\mathbf{x}})$.
		By averaging $Z[\boldsymbol{\psi}](\mathbf{J})$ over realisations of $\mathbf{J}$, we get the effective action.
		This procedure only affects disorder-dependent terms, leading to
         \begin{widetext}
			\begin{equation}
				\begin{aligned}
					\langle \exp{A_1}\rangle_{J} &=\exp\left\{-i \sum_{i,j} \int dt \left[\frac{\mu}{S}\hat{x}_i(t)x_i(t)x_j(t)\right] - \frac{1}{2}\sum_{i,j}\int dt\,dt'\left[\frac{\sigma^2}{S} \hat{x}_i(t)\hat{x}_i(t')x_i(t)x_i(t')x_j(t)x_j(t')\right]\right\}\\
					&\qquad\times\exp\left\{-\frac{1}{2V}\sum_{i,j}\int dt\left[\sqrt{\frac{2}{\pi}}\frac{\sigma}{\sqrt{S}}\hat{x}_i(t)^2 x_i(t) x_j(t) + O(S^{-3/2},V^{-2}) \right]\right\}.
				\end{aligned}
			\end{equation}  
		\end{widetext}
		By introducing the macroscopic order parameters $m(t) = S^{-1}\sum_i x_i(t)$ and $\Delta_2(t,t') = S^{-1}\sum_i x_i(t)x_i(t')$ in the limit $S\rightarrow \infty$, the effective action can be expressed as a sum of $S$ decoupled, identical actions, enabling the description of the system via a single effective stochastic process 		\begin{equation}
			\begin{aligned}
				\dot{x} &= x\left[1-x + \mu m(t) + \sigma\eta(t)\right]\\
				&\qquad+ \sqrt{\frac{x}{V}\left[1+x+\sqrt{\frac{2S}{\pi}}\sigma m(t)\right]}\xi(t),\label{Seq:effproc}
			\end{aligned}
		\end{equation}
		where $\eta(t)$ is a self-consistent noise with the correlator $\langle \eta(t)\eta(t')\rangle = \Delta_2(t,t')$, and we find that $\Delta_2(t,t') \to \langle x(t)x(t') \rangle$ and $m(t) \to \langle x(t)\rangle$, where here $\langle \cdot \rangle$ indicates an average over realisations of the noise $\eta(t)$ and $\xi(t)$. Again, the full details of the saddle-point approximation used to obtain Eq.~(\ref{Seq:effproc}) are given in the Supplementary Material.
        
		In the previously introduced limit $S,V\gg 1$, the term $\sqrt{2S/\pi}\sigma m x /V$ dominates the amplitude of the white noise.
		To ensure that this intrinsic noise remains well-defined and finite, we impose the condition $D\equiv \sqrt{S}/V < \infty$, under which the deterministic limit is now recovered as $D\rightarrow 0$.
		Following this procedure, we arrive at the final expression of the effective dynamics presented in Eq.~(\ref{eq:dmft}).
		
		\begin{figure*}[t!]
			\centering
			\includegraphics[width=\linewidth]{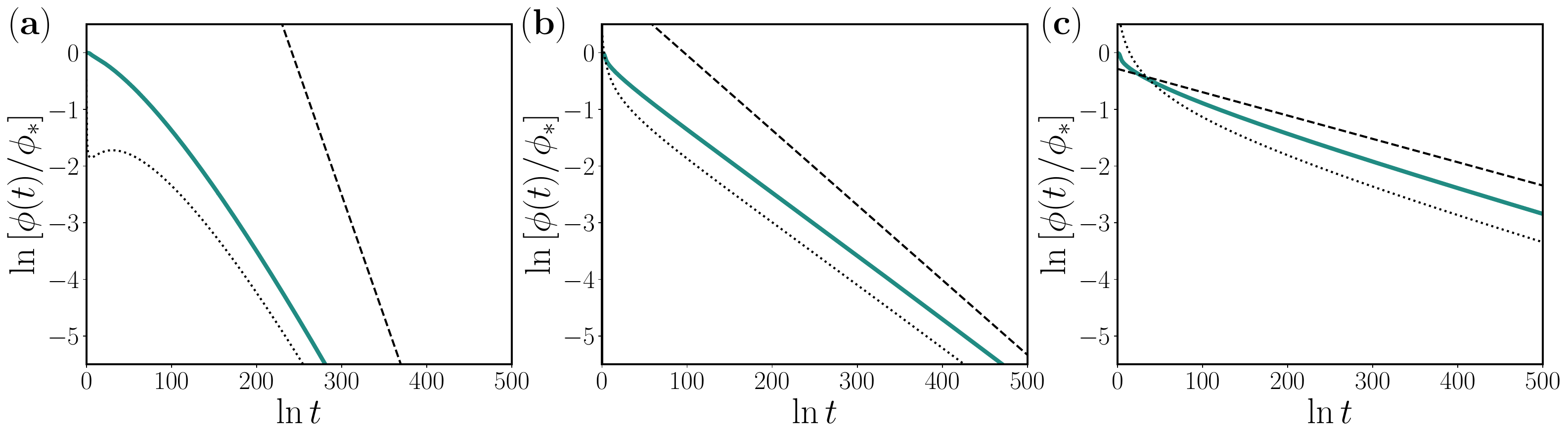}
			\caption{\textbf{Survival probabilities and asymptotes with corrections.} Survival probabilities are shown for different parameter sets: \textbf{(a)} $(\mu,\sigma) = (-1.6,0.4)$, \textbf{(b)} $(\mu,\sigma)=(-0.6,0.8)$, and \textbf{(c)} $(\mu,\sigma)=(0.4,1.2)$. Solid green lines represents numerical solutions derived from the adiabatic theory. Dashed lines denote power-law asymptotes with the leading-order decay exponent $\zeta$ from Eq.~(\ref{eq:decay}), while dotted lines including subleading corrections within $\zeta(t)$.}
			\label{fig:fig4}
		\end{figure*}
		
		\subsection{Post-stationary extinction dynamics}
		Due to stochastic nature of our model, every species inherently faces a finite probability of extinction at any given time.
		In contrast to the deterministic models, this implies that systems will eventually end up with complete extinction ($x_i = 0$ for all $i$).
		However, in the small-noise limit $(D\ll 1)$, the system is expected to persist in the vicinity of the deterministic fixed point for a prolonged duration.
		Accordingly, we consider timescales significantly larger than the relaxation time of the deterministic limit.
		Within this time window, neglecting noise effects, the species abundance distribution is expected to follow the well-known truncated Gaussian~\cite{bunin2017ecological}
		\begin{equation}
			\begin{aligned}
				\rho_*(x) &= [1-\phi_*]\delta(x) + \Theta(x)e^{-A_*(x)},\\
				A_*(x) &= \frac{(x-1-\mu m_*)^2}{2\sigma^2 q_*} + \frac{1}{2} \ln{(2\pi\sigma^2q_*)},
				\label{Seq:ssad}
			\end{aligned}    
		\end{equation}
		where the order parameters $\phi_* = \langle \Theta(x) \rangle_*$, $m_* = \langle x \rangle_*$, and $q_* = \langle x^2 \rangle_*$ come from averaging at the deterministic fixed point.
		
		We now incorporate the effects of demographic noise.
		Under the adiabatic approximation, where the self-consistent parameters $m(t)$ and $\eta(t)$ are treated as static, we write $1+\mu m(t) + \sigma\eta(t) \approx x_*$. 
		The Fokker-Planck equation corresponding to the effective dynamics in Eq.~(\ref{eq:dmft}) is given by
		\begin{align}
			\partial_t \rho(x,t\,|\,x_*,0) = -\partial_x\left[x(x_*-x)\rho\right] + \partial_x^2\left[\lambda_*D x \rho\right],
			\label{Seq:FP}
		\end{align}
		where $\lambda_* = \sigma m_*/\sqrt{2\pi}$.
		This equation describes the evolution of the conditional probability distribution for a given initial value $x_*$, which is itself a random variable drawn from the stationary abundance distribution $\rho_*$ displayed in Eq.~(\ref{Seq:ssad}).
		To solve this, we adopt the WKB ansatz $\rho(x,t\,|\,x_*,0) = \mathcal{C}(x_*)e^{-t/\tau_e(x_*)}e^{-A(x\,|\,x_*)/D}$.
		Assuming that the mean extinction time $\tau_e$ is exponentially large so that $\tau_e^{-1} \approx 0$, we substitute this ansatz into Eq.~(\ref{Seq:FP}).
		Collecting terms of $O(D^{-1})$ yields the action along the extinction trajectory
		\begin{align}
			A(x\,|\,x_*) = \frac{x(x-2x_*)}{2\lambda_*}.
		\end{align} 
		By adjusting the normalisation factor $\mathcal{C}(x_*)$, the conditional probability $\rho(x,t\,|\,x_*,0)$ is given by a Gaussian distribution with mean $x_*$ and variance $\lambda_* D$.
		The post-stationary abundance distribution $\rho(x,t)$ is then obtained by marginalising over $x_*$, leading to
		\begin{equation}
			\begin{aligned}
				\rho(x,t) &\approx \int dx_*\,\rho(x,t\,|\,x_*,0)\rho_*(x_*)\\
				&= [1-\phi(t)]\delta(x) + \Theta(x)\int dx_*\,e^{-A(x,x_*,t)},\\
				A(x,x_*,t) &= A_*(x_*) + \frac{t}{\tau(x_*)} + \frac{(x-x_*)^2}{2\lambda_* D}+ \frac{1}{2}\ln{(2\pi\lambda_* D)}.
				\label{Seq:qsad}
			\end{aligned}    
		\end{equation}
		In the small-noise limit $D\ll 1$, applying the saddle-point approximation to the integral over $x_*$ yields the expression in Eq.~(\ref{eq:qsad}).
		
		The mean extinction time $\tau_e$ is equivalent to the mean first passage time (MFPT) to the absorbing boundary at $x=0$.
		Starting from the adiabatic Fokker-Planck equation in Eq.~(\ref{Seq:FP}), the MFPT $\tau_e(y)$ satisfies following adjoint differential equation
		\begin{align}
			-1 = y(x_*-y)\tau_e' + \lambda_* D y \tau_e'',
			\label{Seq:MFPT}
		\end{align}
		where $y$ denotes the initial abundance.
		In the spirit of the WKB approximation, the conditional abundance distributions are sharply peaked at $x_*$.
		We thus seek the mean extinction time starting from this peak, which is given by
		\begin{equation}
			\begin{aligned}
				\tau_e(x_*) &= \frac{1}{\lambda_* D}\int_0^{x_*}  du\exp\left\{\frac{1}{\lambda_* D} \left(\frac{1}{2}u^2 - x_* u)\right)\right\}\\
				&\qquad\times\int_u^\infty ds\,s^{-1}\exp\left\{-\frac{1}{\lambda_* D}\left(\frac{1}{2}s^2 - x_* s\right)\right\}
			\end{aligned}
		\end{equation}
		Since $u<x_*$, the integration interval for the inner integral $[u,\infty]$ always embraces the sharp peak of integrand at $s=x_*$.
		Evaluating this integral via the saddle-point approximation leads to the solution presented in Eq.~(\ref{eq:met}).
		
		\subsection{Heavy-tailed decay of survival probability}
		The mean extinction time is fundamentally governed by the stationary abundance.
		Based on Eq.~(\ref{eq:met}), we identify the following asymptotic behaviour
		\begin{equation}
			\begin{aligned}
				\tau_e(x) \approx
				\begin{cases}
					\displaystyle
					\sqrt{\frac{2\pi}{\lambda_*D}} &\text{for}~x\ll x_c\\
					\\
					\displaystyle
					\frac{\sqrt{2\pi\lambda_*D}}{x^2}\exp\left(\frac{x^2}{2\lambda_* D}\right)&\text{for}~x\gg x_c
				\end{cases},
			\end{aligned}
		\end{equation}
		where $x_c = \sqrt{2\lambda_* D}$ behaves as a characteristic crossover point.
		This bifurcated behaviour of $\tau_e(x)$ suggests that species with low and high abundances manifest distinct extinction dynamics.
		
		At early times $t\ll 1$, extinction is dominated by low-abundance species.
		As their extinction times are constant independent of their abundances, the survival probability exhibits linear decay
		\begin{align}
			\phi(t) \approx \phi_*\left[1-\epsilon\left(1 - e^{-\sqrt{\frac{\lambda_* D}{2\pi}}t}\right)\right]\approx \phi_*\left(1 - \sqrt{\frac{\lambda_* D}{2\pi}}\epsilon t \right),
		\end{align}
		where the small parameter $\epsilon = \phi_*^{-1}\int_{0^+}^{x_c}dx\,\rho_*(x)$ represents the fraction of the low-abundance species.
		Conversely, in the long-time limit $t \gg 1$, only high-abundance species persist in the remaining community.
		Thus, we can approximate the survival probability as
		\begin{align}
			\phi(t) \approx \int_{x_c}^\infty dx \exp\left\{-\frac{x^2 t}{\sqrt{2\pi \lambda_* D}}e^{-\frac{x^2}{2\lambda_* D}} - A_*(x)\right\},
		\end{align}
        where $A_*$ is the action of the deterministic fixed point defined in Eq.~(\ref{Seq:ssad}).
        For sufficiently large $t$, we can employ the saddle-point approximation with respect to $x$.
		There exists a nonzero local minimum of the action for large times $t>e\sqrt{2\pi\lambda_* D}/(2\sigma^2 q_*)$, located at
		\begin{equation}
			\begin{aligned}
				x_\mathrm{min} &= \sqrt{-2\lambda_* D W_{-1}\left(-\frac{\sqrt{2\pi\lambda_* D}}{2\sigma^2 q_* t}\right)} \\
				&= \sqrt{ 2\lambda_* D\left[ \ln{\left(\frac{2\sigma^2 q_* t}{\sqrt{2\pi\lambda_* D}}\right)} + \ln{\ln{\left(\frac{2\sigma^2 q_* t}{\sqrt{2\pi\lambda_* D}}\right)}} + \cdots\right]},
			\end{aligned}
		\end{equation}
		where $W_{-1}(x)$ is the lower branch of the Lambert $W$ function.
		This saddle point dominates the integral, yielding a long-term asymptote characterised by an anomalous heavy-tailed decay, $\phi(\bar{t}) \propto \bar{t}^{-\zeta(\bar{t})}$ with the time-dependent exponent
		\begin{equation}
			\begin{aligned}
				\zeta(t) \approx \frac{\lambda_* D}{\sigma^2 q_*}  - \frac{1 + \mu m_*}{\sigma^2 q_*} \sqrt{\frac{2\lambda_* D}{\ln{\bar{t}}}} +\left(\frac{\lambda_* D}{\sigma^2 q_*}  + \frac{1}{2} \right) \frac{\ln{\ln{\bar{t}}}}{\ln{\bar{t}}},
			\end{aligned}
		\end{equation}
		where $\bar{t} = 2\sigma^2 q_* t/\sqrt{2\pi\lambda_* D}$ is the rescaled time.
		The constant decay exponent is recovered in the limit $\zeta = \lim_{\bar{t}\rightarrow\infty} \zeta(\bar{t})$, consistent with Eq.~(\ref{eq:decay}).
		Notably, these time-dependent corrections become crucial for large, yet not exponentially large timescales (Fig.~\ref{fig:fig4}). 
		
		Finally, we emphasise that these results hold under the adiabatic approximation, where $m(t) \approx m_*$ and $q(t) \approx q_*$.
		On exponentially large timescales, however, the temporal evolution of these order parameters becomes non-negligible.
		As $m(t)$ gradually decreases over time and the noise amplitude scales with $\sqrt{m(t)}$, we anticipate that the original rule-based model will exhibit an even slower rate of extinction than predicted by the adiabatic theory.
		
		\begin{acknowledgments}
			JWB thanks the Leverhulme Trust for support through the Leverhulme Early Career Fellowship scheme.
		\end{acknowledgments}
		
	\end{CJK}
    \endgroup

\begingroup
\def\addcontentsline#1#2#3{}%
\endgroup

\clearpage

\renewcommand{\thesection}{S\arabic{section}}
\renewcommand{\thefigure}{S\arabic{figure}}
\renewcommand{\theequation}{S\arabic{equation}}

\renewcommand{\citenumfont}[1]{S#1}
\renewcommand{\bibnumfmt}[1]{[S#1]}

\setcounter{section}{0}
\setcounter{figure}{0}
\setcounter{equation}{0}
\setcounter{affil}{0}

\title{Extinction drives emergent metastability in complex ecosystems\\
~\vspace{-0.5em}\\
\Large --- Supplementary Information ---}

\author{Jong Il Park}
\affiliation{Department of Mathematical Sciences, University of Bath, Bath, BA2 7AY, United Kingdom}
\author{Tim Rogers}
\email[Corresponding author: ]{ma3tcr@bath.ac.uk}
\affiliation{Department of Mathematical Sciences, University of Bath, Bath, BA2 7AY, United Kingdom}
\author{Joseph W. Baron}
\email[Corresponding author: ]{jwb96@bath.ac.uk}
\affiliation{Department of Mathematical Sciences, University of Bath, Bath, BA2 7AY, United Kingdom}

\maketitle

\onecolumngrid
\tableofcontents
\clearpage

\section{Complex birth-death systems}
This Supplemental file fully details how we derive the adiabatic approximation for the species abundance distribution and the time-dependent fraction of surviving species, beginning with the rule-based process. We also discuss the various numerical methods used to produce the figures of the main text.
\subsection{Reaction rules and the chemical master equation}
We start with a parsimonious set of pairwise reaction rules, which are
\begin{equation}
\begin{aligned}
    &X_i \xrightarrow{\lambda_b}2X_i,\quad X_i \xrightarrow{\lambda_d} \emptyset &~\text{(intrinsic birth/death)},\\
    &2X_i \xrightarrow{1/V} X_i&~\text{(self-regulation)},\\
    &X_i+X_j \xrightarrow{J_{ij}^-/V}X_j&~\text{(amensalism)},\\
    &X_i+X_j\xrightarrow{J_{ij}^+/V} 2X_i+X_j&~\text{(commensalism)}.
    \label{Seq:rbm}
\end{aligned}
\end{equation}
Here, $V$ is a characteristic system size.
In the well-mixed (spatially homogeneous) case, population state $\mathbf{n}$ is governed by the master equation given by
\begin{equation}
\begin{aligned}
    \partial_t P(\mathbf{n},t) = \sum_i \left[ (\mathcal{E}_i^- - 1) W^{+}(\mathbf{n}) + (\mathcal{E}_i^+ - 1)W^-(\mathbf{n})\right]P(\mathbf{n},t)
\end{aligned}    
\end{equation}
with transition rates
\begin{equation}
	\begin{aligned}
		W^+(\mathbf{n}) &= \lambda_b n_i + \sum_{j\in \partial i^+}J_{ij}^+ \frac{n_i n_j}{V},\\
		W^-(\mathbf{n}) &= \lambda_d n_i + \frac{n_i(n_i -1)}{V} + \sum_{j\in \partial i^-}J_{ij}^- \frac{n_i n_j}{V},
	\end{aligned}    
\end{equation}
where $\mathcal{E}_i^{\pm}:n_i\mapsto n_i\pm 1$ denote the step operators, and $\partial i^{\pm}$ represent disjoint sets of neighbours of species $i$ determined by the assigned interactions $J^{\pm}_{ij}$.

\subsection{Expansion to the approximate Langevin equation}
For large $V$, we may write $n_i(t) = Vq_i(t) + \sqrt{V}\zeta_i(t)$.
We transform $P(\mathbf{n},t)$ to $\Pi (\boldsymbol{\zeta},t)$ via the change of variables~\cite{Svankampen2007stochastic,Sgardiner2009stochastic}, where the transformed density satisfies
\begin{align}
    \partial_t P(\mathbf{n},t) = d_t P(\mathbf{n},t) = \partial_t \Pi(\boldsymbol{\zeta},t) - d_t\mathbf{q}(t) \cdot V^{1/2} \partial_{\boldsymbol{\zeta}}\Pi(\boldsymbol{\zeta},t).
\end{align}
Expanding the step operators as $\mathcal{E}_i^\pm = 1 \pm V^{-1/2}\partial_{\zeta_i} + \frac{1}{2} V^{-1}\partial_{\zeta_i}^2 + O (V^{-3/2})$ and substituting into the master equation gives
\begin{equation}
\begin{aligned}
    \partial_t \Pi - V^{1/2} d_t \mathbf{q} \cdot \partial_{\boldsymbol{\zeta}}\Pi = &-V^{1/2} \sum_i \left[q_i(1 - q_i)\partial_{\zeta_i} +\left( \sum_{j\in\partial i^+} J_{ij}^+ - \sum_{j \in \partial i^-} J_{ij}^-\right)q_iq_j\partial_{\zeta_i}\right]\Pi\\
    &+\frac{1}{2}\sum_i\left[q_i(1+
    q_i) \partial_{\zeta_i}^2 + \left( \sum_{j\in\partial i^+} J_{ij}^+ + \sum_{j\in\partial i^-} J_{ij}^-\right)q_iq_j\partial_{\zeta_i}^2\right]\Pi\\
    &-\sum_i\left[(1-2q_i)\partial_{\zeta_i}\zeta_i + \left( \sum_{j\in\partial i^+} J_{ij}^+ - \sum_{j\in\partial i^-} J_{ij}^-\right)\partial_{\zeta_i}(q_i\zeta_j + q_j\zeta_i)\right]\Pi.
\end{aligned}    
\end{equation}
We introduce the interaction matrix $J_{ij}$, from which we define $J_{ij}^{\pm} = \max(0,\pm J_{ij})$, ensuring $J_{ij} = J_{ij}^+ - J_{ij}^-$ and $|J_{ij}| = J_{ij}^+ + J_{ij}^-$.
Collecting the terms of $O(V^{1/2})$, we obtain the deterministic macroscopic dynamics
\begin{align}
    \dot{q}_i(t) = q_i(t)\left[1-q_i(t) +\sum_{j\setminus i} J_{ij} q_j(t)\right].
\end{align}
We get the linear Fokker-Planck equation for the fluctuation $\zeta_i$ by collecting the terms of $O(V^0)$, and thus the corresponding Langevin equation is given by
\begin{align}
    \dot{\zeta}_i(t) &= \zeta_i(t) \left[1 - 2q_i(t) + \sum_{j\setminus i} J_{ij} q_j(t)\right] + q_i(t) \sum_{j\setminus i} J_{ij}\zeta_j(t) + \sqrt{q_i(t)\left(1 + q_i(t) + \sum_{j\setminus i} |J_{ij}| q_j(t) \right)}\xi_i(t),
    \label{Seq:grlv}
\end{align}
where $\xi_i(t)$ is a Gaussian white noise satisfying ${\langle \xi_i (t)\rangle}_\xi = 0$ and ${\langle \xi_i(t)\xi_j(t')\rangle}_\xi = \delta_{ij}\delta(t-t')$.
We finally get an approximate Langevin equations for the abundance $x_i(t) \equiv n_i/V = q_i(t) + V^{-1/2}\zeta_i(t)$ up to $O(V^{-1/2})$, given by
\begin{align}
    \dot{x}_i(t) = x_i(t) \left[1 - x_i(t) + \sum_{j\setminus i} J_{ij} x_j(t) \right] + V^{-1/2}\sqrt{x_i(t)\left[1+x_i(t)+\sum_{j\setminus i}|J_{ij}|x_j(t)\right]}\xi_i(t),
    \label{Seq:kurtz}
\end{align}
which corresponds to the diffusion approximation rigorously justified by the Kurtz theorem~\cite{Skurtz1978strong,Sgillespie2000chemical}.

\subsection{Dynamical mean-field theory (DMFT)}
To obtain the mean-field dynamics of the system, we employ the Martin-Siggia-Rose-De Dominicis-Janssen (MSRDJ) formalism~\cite{Sgalla2024generating,Sbaron2026lecture}.
We begin by constructing the moment-generating functional (MGF) corresponding to the dynamics in Eq.~(\ref{Seq:kurtz}) as
\begin{equation}
\begin{aligned}
    Z[\boldsymbol{\psi}](\mathbf{J}) &= \int \mathcal{D}[\mathbf{x},i\hat{\mathbf{x}}] \exp\left\{\sum_i\int dt\,x_i(t)\psi_i(t)\right\}\\
    &\hspace{6em}\times\exp \left\{\sum_i\int dt\: \hat{x}_i(t)\left[\dot{x}_i(t) -x_i(t) \left(1 - x_i(t) +\sum_{j\setminus i}J_{ij}x_j(t)\right)\right.\right.\\
    &\hspace{17em}\left.\left.+\frac{1}{2V}\hat{x}_i(t)x_i(t)\left(1+x_i(t) + \sum_{j\setminus i} |J_{ij}|x_j(t)\right)\right]\right\}\\
    & \equiv \int \mathcal{D}[\mathbf{x},i\hat{\mathbf{x}}] \:\exp \left\{\mathcal{A}_0[\mathbf{x},\hat{\mathbf{x}}] + \mathcal{A}_\mathrm{dis}[\mathbf{x},\hat{\mathbf{x}}](\mathbf{J}) + \int dt\, \mathbf{x}(t)\cdot\boldsymbol{\psi}(t)\right\},
\end{aligned}
\label{Seq:mgf}
\end{equation}
where $\int \mathcal{D}[\mathbf{x},i\mathbf{\hat{x}}] = \int_{-\infty}^\infty \int_{-i\infty}^{i\infty}\prod_t [d\mathbf{x}(t)\:d\hat{\mathbf{x}}(t)/(2\pi i)]$ is a functional integral.
We separate the action into two parts, $\mathcal{A}_0$ and $\mathcal{A}_\mathrm{dis}$, where only $\mathcal{A}_\mathrm{dis}$ depends on the disorder $\mathbf{J}$. 
We next take an average over the realisations of $\mathbf{J}$, then for i.i.d. random variables $J_{ij}$ we get
\begin{equation}
\begin{aligned}
    \left\langle \exp \mathcal{A}_\mathrm{dis}(\mathbf{J})\right\rangle_J &= \left\langle \exp \left\{\sum_{i\neq j} \int dt \left[-J_{ij}\hat{x}_i(t)x_i(t) x_j(t) + \frac{|J_{ij}|}{2V}\hat{x}_i(t)\hat{x}_i(t)x_i(t)x_j(t)\right]\right\}\right\rangle_J\\
    &= \exp \left\{\sum_{i\neq j}\int dt \left[-{\langle J_{ij}\rangle}_{J}\hat{x}_i(t)x_i(t)x_j(t)+\frac{{\langle |J_{ij}|\rangle}_J}{2V}\hat{x}_i(t)\hat{x}_i(t)x_i(t)x_j(t)\right]\right.\\
    &\hspace{3.5em}\left.+\frac{1}{2}\sum_{i\neq j}\int dt\:dt'\:\left[\vphantom{\frac{1}{2}}{\llangle J_{ij}^2\rrangle}_J\hat{x}_i(t)x_i(t)x_j(t)\hat{x}_i(t')x_i(t')x_j(t') \right. \right.\\
    &\hspace{10.5em}\left.\left. + \frac{{\llangle J_{ij} |J_{ij}| \rrangle}_J}{2V} \left(\hat{x}_i(t)x_i(t)x_j(t)\hat{x}_i(t')\hat{x}_i(t')x_i(t')x_j(t') + \cdots\right)\right.\right.\\
    &\hspace{10.5em}\left.\left.+ \frac{{\llangle |J_{ij}|^2\rrangle}_J}{4V^2}\hat{x}_i(t)\hat{x}_i(t)x_i(t)x_j(t)\hat{x}_i(t')\hat{x}_i(t')x_i(t')x_j(t')\right] + \cdots\right\},
\end{aligned}
\end{equation}
where we use $\llangle f^n g^m \rrangle$ to indicate the $(n,m)$-th mixed cumulant of $f$ and $g$.
We introduce macroscopic $n$-point correlators as 
\begin{align}
    \Delta_n(t_1,\cdots,t_n) = S^{-1}\sum_{i}x_i(t_1)\,\cdots\,x_i(t_n)~\text{and}~m(t) = \Delta_1(t) = S^{-1}\sum_i x_i(t),
\end{align}
and impose these constraints into the action via delta functionals.
If $(n,m)$-th mixed culmulants of $J_{ij}$ and $|J_{ij}|$ can be written in a form of $\llangle J_{ij}^n |J_{ij}|^m \rrangle = \kappa_{nm}/S$, it is available to perform the saddle-point approximation for the functional integral in Eq.~(\ref{Seq:mgf}).
By this procedure, we get the disorder-averaged MGF as 
\begin{align}
    Z_{\mathrm{eff}} = {\langle Z[\boldsymbol{\psi}](\mathbf{J})\rangle}_J = \int \mathcal{D}[\mathbf{x},i\hat{\mathbf{x}}]\exp\left\{\mathcal{A}_{\mathrm{eff}}[\mathbf{x},\hat{\mathbf{x}}] + \int dt\,\mathbf{x}\cdot\boldsymbol{\psi}\right\}
\end{align}
with an effective action
\begin{equation}
\begin{aligned}
    \mathcal{A}_\mathrm{eff} = \mathcal{A}_0 + \sum_{i}&\left\{-\int dt\,\hat{x}_i(t)\left[\kappa_{10}x_i(t)m(t)\right]\right. \\&\qquad+ \left.\int dt\,dt'\hat{x}_i(t)\hat{x}_i(t')\left[\frac{\kappa_{20}}{2}x_i(t)x_i(t')\Delta_2(t,t') + \frac{\kappa_{01}}{2V}\delta(t-t')x_i(t)m(t)\right] + \cdots\right\},
\end{aligned}    
\end{equation}
where the correlators are now self-consistently defined
\begin{align}
    \Delta_n(t_1,\cdots,t_n) = S^{-1}\sum_i {\langle x_i(t_1)\,\cdots x_i(t_n)\rangle}_\mathrm{eff}\quad\mathrm{with}\quad {\langle \cdots \rangle}_{\mathrm{eff}} = \frac{1}{Z_{\mathrm{eff}}[\mathbf{0}]} \int \mathcal{D}[\mathbf{x},i\hat{\mathbf{x}}]\,(\cdots)e^{\mathcal{A}_\mathrm{eff}}.
\end{align}
The effective action is decoupled, thus an effective single-species dynamics generated by $Z_\mathrm{eff}$ is given by
\begin{align}
 \dot{x}(t) = x(t) \left[1 - x(t)+ \kappa_{10}m(t) + \eta(t)\right] + \Xi(t),
\end{align}
where $\eta(t)$ and $\Xi(t)$ are the zero-mean uncorrelated noises with correlators
\begin{align}
    {\llangle \eta(t_1)\,\cdots\,\eta(t_n)\rrangle}_{\eta} = \kappa_{n0} \Delta_n(t_1,\cdots,t_n)\quad\mathrm{for}\quad n\geq2
\end{align}
and
\begin{equation}
\begin{aligned}
    {\llangle \Xi(t_1)\,\cdots\,\Xi(t_n)\rrangle}_\Xi &= x(t)\left[1+x(t)\right]\delta_{n,2} \\
    &\qquad+ \sum_{k=1}^{\lfloor{n/2}\rfloor}\frac{n!}{(n-2k)!k!} \frac{\kappa_{n-2k,k}}{2^kV^k} x(t_1)\,\cdots \,x(t_{n-k})\,\delta(t_1-t_{n-k+1})\,\cdots\,\delta(t_1-t_n)\Delta_{n-k},
\end{aligned}
\end{equation}
respectively.

\subsection{Gaussian ensemble and dilute system}
Consider a new random matrix $W_{ij} \equiv A_{ij}J_{ij}$, where $A_{ij}=A_{ji}\sim\mathrm{Bernoulli}(K/S)$ encodes the interaction structure and $J_{ij}\sim\mathcal{N}(\mu/K,\sigma^2/K)$ represents the interaction strength.
Equivalently, the joint distribution of $(W_{ij},W_{ji})$ is
\begin{equation}
\begin{aligned}
    P(W_{ij},W_{ji}) &= \left(1-\frac{K}{S}\right)\delta(W_{ij})\delta(W_{ji}) + \frac{K}{S}\Pi(W_{ij},W_{ji}),\\
    \Pi(W_{ij},W_{ji}) &= \frac{1}{2\pi\sigma^2/K}\exp\left\{-\frac{(W_{ij}-\mu/K)^2}{2\sigma^2/K}-\frac{(W_{ji}-\mu/K)^2}{2\sigma^2/K}\right\}.
\end{aligned}    
\end{equation}
The parameter $K$ denotes an average number of interaction per species.
To compute the cumulants $\kappa_{nm}$ of $\mathbf{W}$, we first examine the statistics of $|J_{ij}|$.
Let $\epsilon = \mu/\sigma\sqrt{K}$, then
\begin{equation}
\begin{aligned}
    {\langle |J_{ij}|\rangle}_J &= \sqrt{\frac{2}{\pi}} \frac{\sigma^2}{\mu}\epsilon e^{-\epsilon^2/2} - \frac{\sigma^2}{\mu}\epsilon^2 \mathrm{erf}\left(-\frac{\epsilon}{\sqrt{2}}\right) = \sqrt{\frac{2}{\pi}} \frac{\sigma^2}{\mu}\epsilon + O(\epsilon^3) = \sqrt{\frac{2}{\pi}}\frac{\sigma}{\sqrt{K}} + O(K^{-3/2}),\\
    {\llangle |J_{ij}|^2\rrangle}_J &= {\langle J_{ij}^2 \rangle}_J - {{\langle |J_{ij}| \rangle}^2_J} = \left(1 - \frac{2}{\pi}\right)\frac{\sigma^2}{K} + O(K^{-2})
\end{aligned}    
\end{equation}
Thus, the cumulants of $\mathbf{W}$ follow $\kappa_{nm} = S{\llangle A_{ij} J_{ij}^n |J_{ij}|^m\rrangle}_{A,J} = K\llangle{J_{ij}^n |J_{ij}|^m\rrangle}_J$, leading to
\begin{equation}
\begin{aligned}
    &\kappa_{10} = \mu,~~\kappa_{20} = \sigma^2,~~\kappa_{n0} = 0~\text{for}~n\geq 3,\\
    &\kappa_{01} = \sigma \sqrt{2K/\pi} + O(K^{-1/2}), ~~\kappa_{02} = \sigma^2(1-2/\pi) + O(K^{-1}), ~~\kappa_{0m} \sim O(K^{1-m/2}),\\
    &\kappa_{nm} \approx K{\langle J_{ij}^n|J_{ij}|^m\rangle}_J\sim O(K^{1-\lfloor (n+1)/2 \rfloor -m/2})~\text{for}~m\geq 1.
\end{aligned}    
\end{equation}
For the dense regime $1 \ll K \ll N$, keeping terms up to $O(K^0,V^{-1})$ gives rise to
\begin{align}
    \dot{x} = x(t)\left[1-x(t)+\mu m(t) + \sigma\eta(t)\right]+\sqrt{V^{-1}x(t)\left[1+x(t)+\sigma\sqrt{2K/\pi}m(t)\right]}\xi(t),
    \label{Seq:dmft}
\end{align}
where $\eta(t)$ is a Gaussian noise with correlator $\Delta_{2}(t,t')$, and $\xi(t)$ is a Gaussian white noise.

\subsection{Post-stationary abundance distribution}
In order to solve Eq.~(\ref{Seq:dmft}), we assume that the demographic noise is sufficiently weak so that the relaxation toward the deterministic equilibrium occurs much faster than the extinction driven by the stochastic demographic fluctuations.
This adiabatic assumption allows us to decompose the dynamics into two-stage process.
The system reaches to the deterministic equilibrium, given by a truncated Gaussian distribution
\begin{equation}
\begin{aligned}
    x_* &= \max(0,1+\mu m_* + \sigma\sqrt{q_*} z)~\text{with}~z\sim\mathcal{N}(0,1),\\
    \rho_*(x_*) &= (1-\phi_*)\delta(x_*) + \frac{\Theta(x_*)}{\sqrt{2\pi\sigma^2 q_*}}\exp\left\{- \frac{(x_* - 1 - \mu m_*)^2}{2\sigma^2 q_*}\right\},
\end{aligned}
\end{equation}
where the self-consistent order parameters are defined as
\begin{align}
    \phi_* = \int_{0^+}^\infty dx\,\rho_*(x),\quad m_* = \int_{0^+}^\infty dx\,x\rho_*(x),\quad\text{and}\quad q_* = \int_{0^+}^\infty dx\,x^2\rho_*(x).
\end{align}

In the subsequent extinction stage, where species with low abundance predominantly die out, we approximate $m(t)\approx m_*$ and $\eta(t) \approx \sqrt{q_*} z$.
Under this approximation, Eq.~(\ref{Seq:dmft}) reduces to the adiabatic dynamics for the surviving species
\begin{align}
    \dot{x} = x(x_* - x) + \sqrt{V^{-1}x(1+x+\sigma\sqrt{2K/\pi}m_*)}\xi(t).
    \label{Seq:adiabatic}
\end{align}
To analyse this extinction process, we try the Wentzel-Krammers-Brillouin (WKB) ansatz $\rho(x,t\,|\,x_*,0) = \mathcal{C}(x_*)e^{-t/\tau_e(x_*)}e^{-A(x\,|\,x_*)/D}$ with $D \equiv \sqrt{K}/V$~\cite{Sassaf2017wkb}. 
For large $K$, the corresponding Fokker-Planck equation associated with Eq.~(\ref{Seq:adiabatic}) becomes
\begin{equation}
\begin{aligned}
    \partial_t \rho(x,t\,|\,x_*,0) = -\partial_x\left[x(x_*-x)\rho(x,t\,|\,x_*,0)\right] + \partial_x^2\left[\lambda_* D x \rho(x,t\,|\,x_*,0)\right],
\end{aligned}
\end{equation}
where $\lambda_* = \sigma m_* / \sqrt{2\pi}$.
Assuming that the extinction time $\tau_e$ is exponentially large so that $\tau_e^{-1}\approx 0$, we substitute the WKB ansatz and collect the leading-order terms of $O(D^{-1})$.
This yield the Hamilton-Jacobi equation
\begin{align}
    H(x,p\,|\,x_*) = xp\left( x_*-x + \lambda_* p\right) = 0,
\end{align}
where $p = \partial_x A(x\,|\,x_*)$ plays the role of an effective momentum.

There are two possible trajectories: $p(x) = 0$ and $p(x) = (x-x_*)/\lambda_*$.
The solution $p(x)=0$ corresponds to the relaxation (deterministic) dynamics, governed by $\dot{x} = \partial_p H = x(x_* - x)$.
The nonzero branch is called activation trajectory, which represents the extinction events induced by demographic noise.
Along this trajectory, the non-trivial action is given by
\begin{align}
    A(x\,|\,x_*) = \int_{x_*}^x du\,p(u) = \frac{(x - x_*)^2}{2\lambda_*}.
\end{align}
Including the normalisation factor $\mathcal{C}$,
we obtain the quasi-stationary abundance distribution in the vicinity of $x_*$
\begin{align}
    \rho_{\mathrm{qs}}(x\,|\,x_*) = e^{t/\tau_e} \rho(x,t\,|\,x_*,0) = \frac{1}{\sqrt{2\pi\lambda_* D}}\exp\left\{-\frac{(x-x_*)^2}{2\lambda_* D}\right\}.
\end{align}
The mean extinction time $\tau_e$ is proportional to the exponential of the action accumulated along the activation trajectory, that is
\begin{align}
    \tau_e(x_*) \simeq \mathcal{B}(x_*)\exp\left\{ \int_{x_*}^0 dx\,p(x)\right\} = \mathcal{B}(x_*)\exp\left\{\frac{x_*^2}{2\lambda_* D}\right\},
\end{align}
where $\mathcal{B}(x_*)$ is a pre-exponential factor.
Although this factor originates from non-WKB contributions, in the present setting we can obtain a more precise estimate for $\tau_e$.
Within the WKB framework, the abundance distribution for fixed $z$ is sharply peaked at $x_*(z)$.
Thus, the mean extinction time starting from $x_*$ is given by
\begin{align}
    \tau_e(x_*) = \frac{1}{\lambda_* D}\int_0^{x_*}  du\exp\left\{\frac{1}{\lambda_* D} \left(\frac{1}{2}u^2 - x_* u)\right)\right\}\int_u^\infty ds\,s^{-1}\exp\left\{-\frac{1}{\lambda_* D}\left(\frac{1}{2}s^2 - x_* s\right)\right\}.
\end{align}
Note that the $u<x_*$ throughout the integration range, so the inner integral over the interval $[u,\infty]$ always contains the sharp peak of the integrand.
Applying the saddle-point approximation, we get
\begin{align}
    \tau_e(x_*) = \sqrt{\frac{2\pi}{\lambda_* D}}\frac{1}{x_*} \exp\left\{\frac{x_*^2}{2\lambda_* D}\right\}\int_0^{x_*}  du\exp\left\{\frac{1}{\lambda_* D} \left(\frac{1}{2}u^2 - x_* u)\right)\right\} = \frac{\pi}{x_*} \mathrm{erfi} \left(\frac{x_*}{\sqrt{2\lambda_* D}}\right),
\end{align}
where $\mathrm{erfi}(x)$ is an imaginary error function.
From the asymptotic expansion of $\mathrm{erfi}(x)$,
\begin{align}
    \tau_e (x_*) = 
    \begin{cases}
        \displaystyle
        \sqrt{\frac{2\pi}{\lambda_* D}}+ O(x_*^2) &\mathrm{for}~x_*\ll x_c\\
        \\
        \displaystyle
        \sqrt{2\pi\lambda_* D}\exp\left\{\frac{x_*^2}{2\lambda_* D}\right\}\left[x_*^{-2} + O(x_*^{-4})\right] &\mathrm{for}~x_*\gg x_c
    \end{cases},
\end{align}
where the cutoff abundance $x_c = \sqrt{2\lambda_* D}$.
For large $x_*$, this expression matches the WKB prediction with the pre-exponential factor $\mathcal{B}(x_*) = x_*^{-2}\sqrt{2\pi\lambda_* D}$.

Finally, we get the post-stationary abundance distribution
\begin{equation}
\begin{aligned}
    \rho(x,t) &= [1-\phi(t)]\delta(x) + \int_0^\infty dx_*\:\rho(x,t\,|\,x_*,0)\rho_*(x_*)\\
    &= [1-\phi(t)]\delta(x) + \int_0^\infty \frac{dx_*}{2\pi\sqrt{\sigma^2 q_* \lambda_* D}}\exp \left\{-\frac{t}{\tau_e(x_*)} - \frac{(x-x_*)^2}{2\lambda_* D} - \frac{(x_* - 1 - \mu m_*)^2}{2\sigma^2 q_*}\right\}\\
    & \approx [1-\phi(t)]\delta(x) + e^{-t/\tau_e(x)}\rho_*(x),
\end{aligned}    
\end{equation}
where the last approximation follows from the fact that the integrand is sharply peaked at $x_* = x$.
The survival probability is then given by
\begin{align}
    \phi(t) = \int_0^\infty dx_*\int_0^\infty dx\: e^{-t/\tau_e(x_*)}\rho_{\mathrm{qs}}(x\,|\,x_*)\rho_*(x_*) = \left\langle{e^{-t/\tau_e}}\right\rangle_*.
\end{align}
In particular, the long-time behaviour of $\phi(t)$ is controlled by the statistics of the inverse extinction time $\tau_e^{-1}$, which may lead to broad, non-exponential decay patterns.

\subsection{Survival probability and heavy-tailed decay}
From the fact that the extinction times of low-abundance species are exponentially smaller than those of high-abundance species, the characteristic extinction timescale strongly depends on the initial abundance at the equilibrium.
For early times $t\ll 1$, extinction events are dominated by species with low abundance $x\ll x_c$, and thus the survival probability exhibits an exponential (approximately linear) decay
\begin{align}
    \phi(t) \simeq \int_{0}^{x_c} dx\,\rho_*(x)e^{-\sqrt{\frac{\lambda_* D}{2\pi}}t} + \int_{x_c}^\infty dx\,\rho_*(x)= \phi_*\left[1- \epsilon\left(1 - e^{-\sqrt{\frac{\lambda_* D}{2\pi}}t}\right)\right]\approx \phi_*\left(1 - \sqrt{\frac{\lambda_* D}{2\pi}} \epsilon t\right),
\end{align}
where the small parameter $\epsilon = \phi_*^{-1}\int_{0^+}^{x_c}dx\,\rho_*(x)$ represents the fraction of the low-abundance species.

For long times $1\ll t <\infty$,  the extinction dynamics are dominated by species with high abundance $x\gg x_c$.
In this regime,
\begin{align}
    \phi(t) \simeq \int_{x_c}^\infty \frac{dx}{\sqrt{2\pi\sigma^2 q_*}}\exp\left\{-\frac{x^2 t}{\sqrt{2\pi \lambda_* D}}e^{-\frac{x^2}{2\lambda_* D}} - \frac{(x-1-\mu m_*)^2}{2\sigma^2 q_*}\right\} \equiv \int_{x_c}^\infty \frac{dx}{\sqrt{2\pi\sigma^2 q_*}} e^{-A(x,t)}.
\end{align}
Since high-abundance species dominate, we assume $A(x,t)$ is sufficiently large to apply a saddle-point approximation.
For $x\gg x_c$,
\begin{align}
    \partial_x A(x,t)&\approx \frac{tx}{\sqrt{2\pi\lambda_* D}}\left[-\frac{x^2}{\lambda_* D}e^{-\frac{x^2}{2\lambda_* D}} + \frac{\sqrt{2\pi\lambda_* D}}{\sigma^2 q_* t}\right],\\
    \partial_x^2 A(x,t)&\approx \frac{t}{\sqrt{2\pi\lambda_* D}}\frac{x^4}{\lambda_*^2D^2}e^{-\frac{x^2}{2\lambda_* D}}.
\end{align}
Solving $\partial_x A(x,t) = 0$ yields three stationary points for $t>e\sqrt{2\pi\lambda_* D}/{(2\sigma^2 q_*)}$: $x=0$ and $x=x_{\pm}$, where $x_{\pm}$ satisfy $ue^{-u} = t_w/t$ with $u = x^2/(2\lambda_* D)$ and $t_w = \sqrt{2\pi\lambda_*D}/(2\sigma^2 q_*)$.
Since $x=0$ is a local minimum, the saddle point is given by the larger root $x_+$, which dominates the integral,
\begin{align}
    x_+ = \sqrt{-2\lambda_* D W_{-1}\left(-\frac{\sqrt{2\pi\lambda_* D}}{2\sigma^2 q_* t}\right)} = \sqrt{ 2\lambda_* D\left[ \ln{\left(\frac{t}{t_w}\right)} + \ln{\ln{\left(\frac{t}{t_w}\right)}} + \cdots\right]},
\end{align}
where $W_{-1}(x)$ is the lower branch of the Lambert $W$ function.
Therefore, the long-time asymptotic behaviour of $\phi(t)$ shows anomalous heavy-tailed decay
\begin{equation}
\begin{aligned}
    \phi(t) \simeq \sqrt{\frac{1}{\sigma^2 q_* \partial_x^2 A(x_+,t)}} e^{-A(x_+,t)} &= \sqrt{\frac{\lambda_* D}{x_+^2}}\exp\left\{-\frac{\lambda_* D}{\sigma^2 q_*} - \frac{(x_+ - 1 - \mu m_*)^2}{2\sigma^2 q_*}\right\}\\
    &\propto (\ln{\bar{t}})^{-1/2}(\bar{t}\ln{\bar{t}})^{-\frac{\lambda_* D}{\sigma^2 q_*}}\exp\left\{\frac{1 + \mu m_*}{\sigma^2 q_*} \sqrt{2 \lambda_* D \ln{\bar{t}}}\right\} = {\bar{t}}^{-\zeta(\bar{t})},
\end{aligned}    
\end{equation}
where the rescaled time $\bar{t} = t/t_w$.
The time-dependent exponent is
\begin{equation}
\begin{aligned}
    \zeta(\bar{t}) \approx \frac{\lambda_* D}{\sigma^2 q_*}  - \frac{1 + \mu m_*}{\sigma^2 q_*} \sqrt{\frac{2\lambda_* D}{\ln{\bar{t}}}} +\left(\frac{\lambda_* D}{\sigma^2 q_*}  + \frac{1}{2} \right) \frac{\ln{\ln{\bar{t}}}}{\ln{\bar{t}}}, 
\end{aligned}
\end{equation}
which eventually approaches the power-law decay with an exponent $\zeta = \lambda_*D/(\sigma^2 q_*)$ for $t \gg t_w$.
The leading exponent is given by the ratio between two self-consistent variances associated with the quenched disorder and the demographic noise.

\section{Numerical simulation}

\subsection{Validation of the approximations via numerics}
\begin{figure}[!t]
    \centering
    \includegraphics[width=0.75\textwidth]{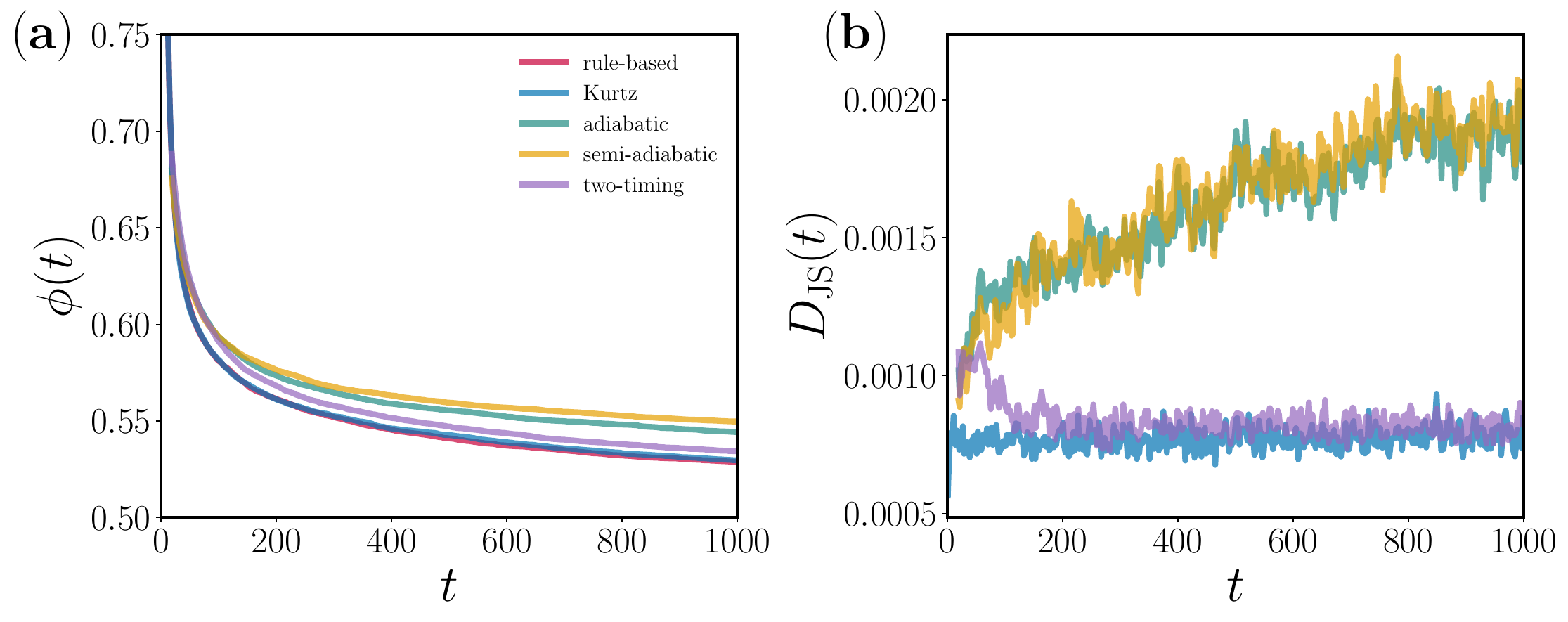}
    \caption{\textbf{Comparison of simulations from various approximations.} \textbf{(a)} Survival probabilities $\phi(t)$ from the rule-based model (red, Eq.~(\ref{Seq:rbm})), approximate Langevin equation (blue, Eq.~(\ref{Seq:kurtz})), adiabatic (green, Eq.~(\ref{Seq:adiabatic})), semi-adiabatic (yellow), and two-timing (purple) approximations. The latter three results are shifted by $t_d = 18.5$. \textbf{(b)} Jensen--Shannon divergence $D_{\mathrm{JS}}(t)$ between rule-based simulation and each of the other approximations. All results use the identical set of 30 independent configurations of interaction matrix $\mathbf{J}$. Parameters: $S=1024, K=1024, V=1000,\mu=-1,\sigma=1$.}
    \label{fig:figS1}
\end{figure}
We numerically examine the validity of the approximations introduced in our theory.
We first assess the accuracy of the approximate Langevin equation in Eq.~(\ref{Seq:kurtz}) by comparing its numerical simulations with those obtained from the rule-based model.
As a quantitative measure of the discrepancy between two descriptions, we compute the Jensen--Shannon divergence
\begin{align}
    D_{\mathrm{JS}}(\rho_1 || \rho_2 ;t) = \frac{1}{2} \int_0^\infty dx\left[ \rho_1(x,t)\ln\frac{\rho_1(x,t)}{\bar{\rho}(x,t)} + \rho_2(x,t) \ln \frac{\rho_2(x,t)}{\bar{\rho}(x,t)}\right]~\mathrm{with}~\bar{\rho} = \frac{\rho_1+\rho_2}{2},
\end{align}
which offers a symmetric measure of the distance between two probability density functions.
By monitoring this divergence together with the survival probability $\phi(t)$, we confirm that, at least for the parameter regimes we investigated, the Langevin approximation provides a reliable description (Fig.~\ref{fig:figS1}). 

To evaluate the validity of the adiabatic approximation in Eq.~(\ref{Seq:adiabatic}), we two variations of it.
In the first, we restore the temporal dependence of order parameters $m(t)$ and $q(t)$, while keep assuming that the correlation time of $\eta(t)$ remains sufficiently large so that $\eta(t) \approx \sqrt{q(t)}z$ (we refer to this as `semi-adiabatic').
In the second variation, we retain the spirit of the adiabatic approach but incorporate full temporal information.
Directly integrating the self-consistent noise is numerically challenging, thus we instead integrate the approximate Langevin equation in Eq.~(\ref{Seq:kurtz}) with initial conditions set to the deterministic equilibrium $\mathbf{x}(0) = \mathbf{x}_*$ (we refer to this as `two-timing').
As shown in Fig.~\ref{fig:figS1}, the two-timing simulations are closely aligned with both the rule-based model and its Langevin approximation, whereas the semi-adiabatic results remain similar to the fully adiabatic one.
These observations indicate that the the correlator $\Delta_2(t,t')$ plays an essential role, and therefore the self-consistent noise $\eta(t)$ cannot be simply reduced to a simple Gaussian random variable.

\subsection{Simulation details}
We first sample a random matrix $\mathbf{J}$ and initialise all species abundances to unity, $\mathbf{x}(0) = [1,\cdots,1]$.
For the deterministic grLV, we numerically integrate the equations in Eq.~(\ref{Seq:grlv}) using the Runge-Kutta-Fehlberg (RK45) method restricting the maximum time step to at $10^{-2}$.
A threshold value $10^{-12}$ is imposed, below which a species abundance is considered extinct.
The Milstein method is used for the stochastic differential equations in Eq.~(\ref{Seq:kurtz}) and Eq.~(\ref{Seq:adiabatic}).
We apply $\tau$-leaping algorithm for simulating the rule-based model in Eq.~(\ref{Seq:rbm}).


\begin{thebibliography}{99}
\makeatletter
\providecommand \@ifxundefined [1]{%
 \@ifx{#1\undefined}
}%
\providecommand \@ifnum [1]{%
 \ifnum #1\expandafter \@firstoftwo
 \else \expandafter \@secondoftwo
 \fi
}%
\providecommand \@ifx [1]{%
 \ifx #1\expandafter \@firstoftwo
 \else \expandafter \@secondoftwo
 \fi
}%
\providecommand \natexlab [1]{#1}%
\providecommand \enquote  [1]{``#1''}%
\providecommand \bibnamefont  [1]{#1}%
\providecommand \bibfnamefont [1]{#1}%
\providecommand \citenamefont [1]{#1}%
\providecommand \href@noop [0]{\@secondoftwo}%
\providecommand \href [0]{\begingroup \@sanitize@url \@href}%
\providecommand \@href[1]{\@@startlink{#1}\@@href}%
\providecommand \@@href[1]{\endgroup#1\@@endlink}%
\providecommand \@sanitize@url [0]{\catcode `\\12\catcode `\$12\catcode `\&12\catcode `\#12\catcode `\^12\catcode `\_12\catcode `\%12\relax}%
\providecommand \@@startlink[1]{}%
\providecommand \@@endlink[0]{}%
\providecommand \url  [0]{\begingroup\@sanitize@url \@url }%
\providecommand \@url [1]{\endgroup\@href {#1}{\urlprefix }}%
\providecommand \urlprefix  [0]{URL }%
\providecommand \Eprint [0]{\href }%
\providecommand \doibase [0]{https://doi.org/}%
\providecommand \selectlanguage [0]{\@gobble}%
\providecommand \bibinfo  [0]{\@secondoftwo}%
\providecommand \bibfield  [0]{\@secondoftwo}%
\providecommand \translation [1]{[#1]}%
\providecommand \BibitemOpen [0]{}%
\providecommand \bibitemStop [0]{}%
\providecommand \bibitemNoStop [0]{.\EOS\space}%
\providecommand \EOS [0]{\spacefactor3000\relax}%
\providecommand \BibitemShut  [1]{\csname bibitem#1\endcsname}%
\let\auto@bib@innerbib\@empty
\bibitem [{\citenamefont {May}(1972)}]{may1972will}%
  \BibitemOpen
  \bibfield  {author} {\bibinfo {author} {\bibfnamefont {R.~M.}\ \bibnamefont {May}},\ }\bibfield  {title} {\bibinfo {title} {Will a {L}arge {C}omplex {S}ystem be {S}table?},\ }\href {https://doi.org/10.1038/238413a0} {\bibfield  {journal} {\bibinfo  {journal} {Nature}\ }\textbf {\bibinfo {volume} {238}},\ \bibinfo {pages} {413} (\bibinfo {year} {1972})}\BibitemShut {NoStop}%
\bibitem [{\citenamefont {Bunin}(2017)}]{bunin2017ecological}%
  \BibitemOpen
  \bibfield  {author} {\bibinfo {author} {\bibfnamefont {G.}~\bibnamefont {Bunin}},\ }\bibfield  {title} {\bibinfo {title} {Ecological communities with {L}otka-{V}olterra dynamics},\ }\href {https://doi.org/10.1103/PhysRevE.95.042414} {\bibfield  {journal} {\bibinfo  {journal} {Phys. Rev. E}\ }\textbf {\bibinfo {volume} {95}},\ \bibinfo {pages} {042414} (\bibinfo {year} {2017})}\BibitemShut {NoStop}%
\bibitem [{\citenamefont {Galla}(2018)}]{galla2018dynamically}%
  \BibitemOpen
  \bibfield  {author} {\bibinfo {author} {\bibfnamefont {T.}~\bibnamefont {Galla}},\ }\bibfield  {title} {\bibinfo {title} {Dynamically evolved community size and stability of random {L}otka-{V}olterra ecosystems},\ }\href {https://doi.org/10.1209/0295-5075/123/48004} {\bibfield  {journal} {\bibinfo  {journal} {Europhys. Lett.}\ }\textbf {\bibinfo {volume} {123}},\ \bibinfo {pages} {48004} (\bibinfo {year} {2018})}\BibitemShut {NoStop}%
\bibitem [{\citenamefont {Ellner}\ and\ \citenamefont {Turchin}(1995)}]{ellner1995chaos}%
  \BibitemOpen
  \bibfield  {author} {\bibinfo {author} {\bibfnamefont {S.}~\bibnamefont {Ellner}}\ and\ \bibinfo {author} {\bibfnamefont {P.}~\bibnamefont {Turchin}},\ }\bibfield  {title} {\bibinfo {title} {Chaos in a {N}oisy {W}orld: {N}ew {M}ethods and {E}vidence from {T}ime-{S}eries {A}nalysis},\ }\href {https://doi.org/10.1086/285744} {\bibfield  {journal} {\bibinfo  {journal} {Am. Nat.}\ }\textbf {\bibinfo {volume} {145}},\ \bibinfo {pages} {343} (\bibinfo {year} {1995})}\BibitemShut {NoStop}%
\bibitem [{\citenamefont {Hastings}\ \emph {et~al.}(1993)\citenamefont {Hastings}, \citenamefont {Hom}, \citenamefont {Ellner}, \citenamefont {Turchin},\ and\ \citenamefont {Godfray}}]{hastings1993chaos}%
  \BibitemOpen
  \bibfield  {author} {\bibinfo {author} {\bibfnamefont {A.}~\bibnamefont {Hastings}}, \bibinfo {author} {\bibfnamefont {C.~L.}\ \bibnamefont {Hom}}, \bibinfo {author} {\bibfnamefont {S.}~\bibnamefont {Ellner}}, \bibinfo {author} {\bibfnamefont {P.}~\bibnamefont {Turchin}},\ and\ \bibinfo {author} {\bibfnamefont {H.~C.~J.}\ \bibnamefont {Godfray}},\ }\bibfield  {title} {\bibinfo {title} {Chaos in {E}cology: {I}s {M}other {N}ature a {S}trange {A}ttractor?},\ }\href {https://www.jstor.org/stable/2097171} {\bibfield  {journal} {\bibinfo  {journal} {Annu. Rev. Ecol. Syst.}\ }\textbf {\bibinfo {volume} {24}},\ \bibinfo {pages} {1} (\bibinfo {year} {1993})}\BibitemShut {NoStop}%
\bibitem [{\citenamefont {Berryman}\ and\ \citenamefont {Millstein}(1989)}]{berryman1989ecological}%
  \BibitemOpen
  \bibfield  {author} {\bibinfo {author} {\bibfnamefont {A.~A.}\ \bibnamefont {Berryman}}\ and\ \bibinfo {author} {\bibfnamefont {J.~A.}\ \bibnamefont {Millstein}},\ }\bibfield  {title} {\bibinfo {title} {Are ecological systems chaotic -- {A}nd if not, why not?},\ }\href {https://doi.org/10.1016/0169-5347(89)90014-1} {\bibfield  {journal} {\bibinfo  {journal} {Trends Ecol. Evol.}\ }\textbf {\bibinfo {volume} {4}},\ \bibinfo {pages} {26} (\bibinfo {year} {1989})}\BibitemShut {NoStop}%
\bibitem [{\citenamefont {Sibly}\ \emph {et~al.}(2007)\citenamefont {Sibly}, \citenamefont {Barker}, \citenamefont {Hone},\ and\ \citenamefont {Pagel}}]{sibly2007stability}%
  \BibitemOpen
  \bibfield  {author} {\bibinfo {author} {\bibfnamefont {R.~M.}\ \bibnamefont {Sibly}}, \bibinfo {author} {\bibfnamefont {D.}~\bibnamefont {Barker}}, \bibinfo {author} {\bibfnamefont {J.}~\bibnamefont {Hone}},\ and\ \bibinfo {author} {\bibfnamefont {M.}~\bibnamefont {Pagel}},\ }\bibfield  {title} {\bibinfo {title} {On the stability of populations of mammals, birds, fish and insects},\ }\href {https://doi.org/10.1111/j.1461-0248.2007.01092.x} {\bibfield  {journal} {\bibinfo  {journal} {Ecol. Lett.}\ }\textbf {\bibinfo {volume} {10}},\ \bibinfo {pages} {970} (\bibinfo {year} {2007})}\BibitemShut {NoStop}%
\bibitem [{\citenamefont {Sugihara}\ \emph {et~al.}(1990)\citenamefont {Sugihara}, \citenamefont {Grenfell},\ and\ \citenamefont {May}}]{sugihara1990distinguishing}%
  \BibitemOpen
  \bibfield  {author} {\bibinfo {author} {\bibfnamefont {G.}~\bibnamefont {Sugihara}}, \bibinfo {author} {\bibfnamefont {B.~T.}\ \bibnamefont {Grenfell}},\ and\ \bibinfo {author} {\bibfnamefont {R.~M.}\ \bibnamefont {May}},\ }\bibfield  {title} {\bibinfo {title} {Distinguishing error from chaos in ecological time series},\ }\href {https://doi.org/10.1098/rstb.1990.0195} {\bibfield  {journal} {\bibinfo  {journal} {Philos. Trans. R. Soc. Lond. B Biol. Sci.}\ }\textbf {\bibinfo {volume} {330}},\ \bibinfo {pages} {235} (\bibinfo {year} {1990})}\BibitemShut {NoStop}%
\bibitem [{\citenamefont {Barahona}\ and\ \citenamefont {Poon}(1996)}]{barahona1996detection}%
  \BibitemOpen
  \bibfield  {author} {\bibinfo {author} {\bibfnamefont {M.}~\bibnamefont {Barahona}}\ and\ \bibinfo {author} {\bibfnamefont {C.-S.}\ \bibnamefont {Poon}},\ }\bibfield  {title} {\bibinfo {title} {Detection of nonlinear dynamics in short, noisy time series},\ }\href {https://doi.org/10.1038/381215a0} {\bibfield  {journal} {\bibinfo  {journal} {Nature}\ }\textbf {\bibinfo {volume} {381}},\ \bibinfo {pages} {215} (\bibinfo {year} {1996})}\BibitemShut {NoStop}%
\bibitem [{\citenamefont {Hunt}\ \emph {et~al.}(2003)\citenamefont {Hunt}, \citenamefont {Antle},\ and\ \citenamefont {Paustian}}]{hunt2003false}%
  \BibitemOpen
  \bibfield  {author} {\bibinfo {author} {\bibfnamefont {H.~W.}\ \bibnamefont {Hunt}}, \bibinfo {author} {\bibfnamefont {J.~M.}\ \bibnamefont {Antle}},\ and\ \bibinfo {author} {\bibfnamefont {K.}~\bibnamefont {Paustian}},\ }\bibfield  {title} {\bibinfo {title} {False determinations of chaos in short noisy time series},\ }\href {https://doi.org/10.1016/S0167-2789(03)00044-7} {\bibfield  {journal} {\bibinfo  {journal} {Physica D}\ }\textbf {\bibinfo {volume} {180}},\ \bibinfo {pages} {115} (\bibinfo {year} {2003})}\BibitemShut {NoStop}%
\bibitem [{\citenamefont {Rogers}\ \emph {et~al.}(2022)\citenamefont {Rogers}, \citenamefont {Johnson},\ and\ \citenamefont {Munch}}]{rogers2022chaos}%
  \BibitemOpen
  \bibfield  {author} {\bibinfo {author} {\bibfnamefont {T.~L.}\ \bibnamefont {Rogers}}, \bibinfo {author} {\bibfnamefont {B.~J.}\ \bibnamefont {Johnson}},\ and\ \bibinfo {author} {\bibfnamefont {S.~B.}\ \bibnamefont {Munch}},\ }\bibfield  {title} {\bibinfo {title} {Chaos is not rare in natural ecosystems},\ }\href {https://doi.org/10.1038/s41559-022-01787-y} {\bibfield  {journal} {\bibinfo  {journal} {Nat. Ecol. Evol.}\ }\textbf {\bibinfo {volume} {6}},\ \bibinfo {pages} {1105} (\bibinfo {year} {2022})}\BibitemShut {NoStop}%
\bibitem [{\citenamefont {Beninc{\`a}}\ \emph {et~al.}(2008)\citenamefont {Beninc{\`a}}, \citenamefont {Huisman}, \citenamefont {Heerkloss}, \citenamefont {J{\"o}hnk}, \citenamefont {Branco}, \citenamefont {Van~Nes}, \citenamefont {Scheffer},\ and\ \citenamefont {Ellner}}]{beninca2008chaos}%
  \BibitemOpen
  \bibfield  {author} {\bibinfo {author} {\bibfnamefont {E.}~\bibnamefont {Beninc{\`a}}}, \bibinfo {author} {\bibfnamefont {J.}~\bibnamefont {Huisman}}, \bibinfo {author} {\bibfnamefont {R.}~\bibnamefont {Heerkloss}}, \bibinfo {author} {\bibfnamefont {K.~D.}\ \bibnamefont {J{\"o}hnk}}, \bibinfo {author} {\bibfnamefont {P.}~\bibnamefont {Branco}}, \bibinfo {author} {\bibfnamefont {E.~H.}\ \bibnamefont {Van~Nes}}, \bibinfo {author} {\bibfnamefont {M.}~\bibnamefont {Scheffer}},\ and\ \bibinfo {author} {\bibfnamefont {S.~P.}\ \bibnamefont {Ellner}},\ }\bibfield  {title} {\bibinfo {title} {Chaos in a long-term experiment with a plankton community},\ }\href {https://doi.org/10.1038/nature06512} {\bibfield  {journal} {\bibinfo  {journal} {Nature}\ }\textbf {\bibinfo {volume} {451}},\ \bibinfo {pages} {822} (\bibinfo {year} {2008})}\BibitemShut {NoStop}%
\bibitem [{\citenamefont {Park}\ \emph {et~al.}(2024)\citenamefont {Park}, \citenamefont {Lee}, \citenamefont {Lee},\ and\ \citenamefont {Park}}]{park2024incorporating}%
  \BibitemOpen
  \bibfield  {author} {\bibinfo {author} {\bibfnamefont {J.~I.}\ \bibnamefont {Park}}, \bibinfo {author} {\bibfnamefont {D.-S.}\ \bibnamefont {Lee}}, \bibinfo {author} {\bibfnamefont {S.~H.}\ \bibnamefont {Lee}},\ and\ \bibinfo {author} {\bibfnamefont {H.~J.}\ \bibnamefont {Park}},\ }\bibfield  {title} {\bibinfo {title} {Incorporating heterogeneous interactions for ecological biodiversity},\ }\href {https://doi.org/10.1103/PhysRevLett.133.198402} {\bibfield  {journal} {\bibinfo  {journal} {Phys. Rev. Lett.}\ }\textbf {\bibinfo {volume} {133}},\ \bibinfo {pages} {198402} (\bibinfo {year} {2024})}\BibitemShut {NoStop}%
\bibitem [{\citenamefont {Poley}\ \emph {et~al.}(2025)\citenamefont {Poley}, \citenamefont {Galla},\ and\ \citenamefont {Baron}}]{poley2025interaction}%
  \BibitemOpen
  \bibfield  {author} {\bibinfo {author} {\bibfnamefont {L.}~\bibnamefont {Poley}}, \bibinfo {author} {\bibfnamefont {T.}~\bibnamefont {Galla}},\ and\ \bibinfo {author} {\bibfnamefont {J.~W.}\ \bibnamefont {Baron}},\ }\bibfield  {title} {\bibinfo {title} {Interaction networks in persistent {L}otka-{V}olterra communities},\ }\href {https://doi.org/10.1103/PhysRevE.111.014318} {\bibfield  {journal} {\bibinfo  {journal} {Phys. Rev. E}\ }\textbf {\bibinfo {volume} {111}},\ \bibinfo {pages} {014318} (\bibinfo {year} {2025})}\BibitemShut {NoStop}%
\bibitem [{\citenamefont {Aguirre-L{\'o}pez}(2024)}]{aguirre2024heterogeneous}%
  \BibitemOpen
  \bibfield  {author} {\bibinfo {author} {\bibfnamefont {F.}~\bibnamefont {Aguirre-L{\'o}pez}},\ }\bibfield  {title} {\bibinfo {title} {Heterogeneous mean-field analysis of the generalized {L}otka--{V}olterra model on a network},\ }\href {https://doi.org/10.1088/1751-8121/ad6ab2} {\bibfield  {journal} {\bibinfo  {journal} {J. Phys. A: Math. Theor.}\ }\textbf {\bibinfo {volume} {57}},\ \bibinfo {pages} {345002} (\bibinfo {year} {2024})}\BibitemShut {NoStop}%
\bibitem [{\citenamefont {Emary}\ and\ \citenamefont {Malchow}(2022)}]{emary2022stability}%
  \BibitemOpen
  \bibfield  {author} {\bibinfo {author} {\bibfnamefont {C.}~\bibnamefont {Emary}}\ and\ \bibinfo {author} {\bibfnamefont {A.-K.}\ \bibnamefont {Malchow}},\ }\bibfield  {title} {\bibinfo {title} {Stability-instability transition in tripartite merged ecological networks},\ }\href {https://doi.org/10.1007/s00285-022-01783-7} {\bibfield  {journal} {\bibinfo  {journal} {J. Math. Biol.}\ }\textbf {\bibinfo {volume} {85}},\ \bibinfo {pages} {20} (\bibinfo {year} {2022})}\BibitemShut {NoStop}%
\bibitem [{\citenamefont {Gravel}\ \emph {et~al.}(2016)\citenamefont {Gravel}, \citenamefont {Massol},\ and\ \citenamefont {Leibold}}]{gravel2016stability}%
  \BibitemOpen
  \bibfield  {author} {\bibinfo {author} {\bibfnamefont {D.}~\bibnamefont {Gravel}}, \bibinfo {author} {\bibfnamefont {F.}~\bibnamefont {Massol}},\ and\ \bibinfo {author} {\bibfnamefont {M.~A.}\ \bibnamefont {Leibold}},\ }\bibfield  {title} {\bibinfo {title} {Stability and complexity in model meta-ecosystems},\ }\href {https://doi.org/10.1038/ncomms12457} {\bibfield  {journal} {\bibinfo  {journal} {Nat. Commun.}\ }\textbf {\bibinfo {volume} {7}},\ \bibinfo {pages} {12457} (\bibinfo {year} {2016})}\BibitemShut {NoStop}%
\bibitem [{\citenamefont {Baron}\ and\ \citenamefont {Galla}(2020)}]{baron2020dispersal}%
  \BibitemOpen
  \bibfield  {author} {\bibinfo {author} {\bibfnamefont {J.~W.}\ \bibnamefont {Baron}}\ and\ \bibinfo {author} {\bibfnamefont {T.}~\bibnamefont {Galla}},\ }\bibfield  {title} {\bibinfo {title} {Dispersal-induced instability in complex ecosystems},\ }\href {https://doi.org/10.1038/s41467-020-19824-4} {\bibfield  {journal} {\bibinfo  {journal} {Nat. Commun.}\ }\textbf {\bibinfo {volume} {11}},\ \bibinfo {pages} {6032} (\bibinfo {year} {2020})}\BibitemShut {NoStop}%
\bibitem [{\citenamefont {Bairey}\ \emph {et~al.}(2016)\citenamefont {Bairey}, \citenamefont {Kelsic},\ and\ \citenamefont {Kishony}}]{bairey2016high}%
  \BibitemOpen
  \bibfield  {author} {\bibinfo {author} {\bibfnamefont {E.}~\bibnamefont {Bairey}}, \bibinfo {author} {\bibfnamefont {E.~D.}\ \bibnamefont {Kelsic}},\ and\ \bibinfo {author} {\bibfnamefont {R.}~\bibnamefont {Kishony}},\ }\bibfield  {title} {\bibinfo {title} {High-order species interactions shape ecosystem diversity},\ }\href {https://doi.org/10.1038/ncomms12285} {\bibfield  {journal} {\bibinfo  {journal} {Nat. Commun.}\ }\textbf {\bibinfo {volume} {7}},\ \bibinfo {pages} {12285} (\bibinfo {year} {2016})}\BibitemShut {NoStop}%
\bibitem [{\citenamefont {Grilli}\ \emph {et~al.}(2017)\citenamefont {Grilli}, \citenamefont {Barab{\'a}s}, \citenamefont {Michalska-Smith},\ and\ \citenamefont {Allesina}}]{grilli2017higher}%
  \BibitemOpen
  \bibfield  {author} {\bibinfo {author} {\bibfnamefont {J.}~\bibnamefont {Grilli}}, \bibinfo {author} {\bibfnamefont {G.}~\bibnamefont {Barab{\'a}s}}, \bibinfo {author} {\bibfnamefont {M.~J.}\ \bibnamefont {Michalska-Smith}},\ and\ \bibinfo {author} {\bibfnamefont {S.}~\bibnamefont {Allesina}},\ }\bibfield  {title} {\bibinfo {title} {Higher-order interactions stabilize dynamics in competitive network models},\ }\href {https://doi.org/10.1038/nature23273} {\bibfield  {journal} {\bibinfo  {journal} {Nature}\ }\textbf {\bibinfo {volume} {548}},\ \bibinfo {pages} {210} (\bibinfo {year} {2017})}\BibitemShut {NoStop}%
\bibitem [{\citenamefont {Hatton}\ \emph {et~al.}(2024)\citenamefont {Hatton}, \citenamefont {Mazzarisi}, \citenamefont {Altieri},\ and\ \citenamefont {Smerlak}}]{hatton2024diversity}%
  \BibitemOpen
  \bibfield  {author} {\bibinfo {author} {\bibfnamefont {I.~A.}\ \bibnamefont {Hatton}}, \bibinfo {author} {\bibfnamefont {O.}~\bibnamefont {Mazzarisi}}, \bibinfo {author} {\bibfnamefont {A.}~\bibnamefont {Altieri}},\ and\ \bibinfo {author} {\bibfnamefont {M.}~\bibnamefont {Smerlak}},\ }\bibfield  {title} {\bibinfo {title} {Diversity begets stability: {S}ublinear growth and competitive coexistence across ecosystems},\ }\href {https://www.science.org/doi/abs/10.1126/science.adg8488} {\bibfield  {journal} {\bibinfo  {journal} {Science}\ }\textbf {\bibinfo {volume} {383}},\ \bibinfo {pages} {eadg8488} (\bibinfo {year} {2024})}\BibitemShut {NoStop}%
\bibitem [{\citenamefont {Neutel}\ and\ \citenamefont {Thorne}(2016)}]{neutel2016linking}%
  \BibitemOpen
  \bibfield  {author} {\bibinfo {author} {\bibfnamefont {A.-M.}\ \bibnamefont {Neutel}}\ and\ \bibinfo {author} {\bibfnamefont {M.~A.~S.}\ \bibnamefont {Thorne}},\ }\bibfield  {title} {\bibinfo {title} {Linking saturation, stability and sustainability in food webs with observed equilibrium structure},\ }\href {https://doi.org/10.1007/s12080-015-0270-z} {\bibfield  {journal} {\bibinfo  {journal} {Theor. Ecol.}\ }\textbf {\bibinfo {volume} {9}},\ \bibinfo {pages} {73} (\bibinfo {year} {2016})}\BibitemShut {NoStop}%
\bibitem [{\citenamefont {D'Odorico}\ \emph {et~al.}(2005)\citenamefont {D'Odorico}, \citenamefont {Laio},\ and\ \citenamefont {Ridolfi}}]{d2005noise}%
  \BibitemOpen
  \bibfield  {author} {\bibinfo {author} {\bibfnamefont {P.}~\bibnamefont {D'Odorico}}, \bibinfo {author} {\bibfnamefont {F.}~\bibnamefont {Laio}},\ and\ \bibinfo {author} {\bibfnamefont {L.}~\bibnamefont {Ridolfi}},\ }\bibfield  {title} {\bibinfo {title} {Noise-induced stability in dryland plant ecosystems},\ }\href {https://doi.org/10.1073/pnas.0502884102} {\bibfield  {journal} {\bibinfo  {journal} {Proc. Natl. Acad. Sci. U.S.A.}\ }\textbf {\bibinfo {volume} {102}},\ \bibinfo {pages} {10819} (\bibinfo {year} {2005})}\BibitemShut {NoStop}%
\bibitem [{\citenamefont {Parker}\ \emph {et~al.}(2011)\citenamefont {Parker}, \citenamefont {Kamenev},\ and\ \citenamefont {Meerson}}]{parker2011noise}%
  \BibitemOpen
  \bibfield  {author} {\bibinfo {author} {\bibfnamefont {M.}~\bibnamefont {Parker}}, \bibinfo {author} {\bibfnamefont {A.}~\bibnamefont {Kamenev}},\ and\ \bibinfo {author} {\bibfnamefont {B.}~\bibnamefont {Meerson}},\ }\bibfield  {title} {\bibinfo {title} {Noise-induced stabilization in population dynamics},\ }\href {https://doi.org/10.1103/PhysRevLett.107.180603} {\bibfield  {journal} {\bibinfo  {journal} {Phys. Rev. Lett.}\ }\textbf {\bibinfo {volume} {107}},\ \bibinfo {pages} {180603} (\bibinfo {year} {2011})}\BibitemShut {NoStop}%
\bibitem [{\citenamefont {Yamazaki}\ \emph {et~al.}(1998)\citenamefont {Yamazaki}, \citenamefont {Yamada},\ and\ \citenamefont {Kai}}]{yamazaki1998can}%
  \BibitemOpen
  \bibfield  {author} {\bibinfo {author} {\bibfnamefont {H.}~\bibnamefont {Yamazaki}}, \bibinfo {author} {\bibfnamefont {T.}~\bibnamefont {Yamada}},\ and\ \bibinfo {author} {\bibfnamefont {S.}~\bibnamefont {Kai}},\ }\bibfield  {title} {\bibinfo {title} {Can stochastic resonance lead to order in chaos?},\ }\href {https://doi.org/10.1103/PhysRevLett.81.4112} {\bibfield  {journal} {\bibinfo  {journal} {Phys. Rev. Lett.}\ }\textbf {\bibinfo {volume} {81}},\ \bibinfo {pages} {4112} (\bibinfo {year} {1998})}\BibitemShut {NoStop}%
\bibitem [{\citenamefont {Altieri}\ \emph {et~al.}(2021)\citenamefont {Altieri}, \citenamefont {Roy}, \citenamefont {Cammarota},\ and\ \citenamefont {Biroli}}]{altieri2021properties}%
  \BibitemOpen
  \bibfield  {author} {\bibinfo {author} {\bibfnamefont {A.}~\bibnamefont {Altieri}}, \bibinfo {author} {\bibfnamefont {F.}~\bibnamefont {Roy}}, \bibinfo {author} {\bibfnamefont {C.}~\bibnamefont {Cammarota}},\ and\ \bibinfo {author} {\bibfnamefont {G.}~\bibnamefont {Biroli}},\ }\bibfield  {title} {\bibinfo {title} {Properties of {E}quilibria and {G}lassy {P}hases of the {R}andom {L}otka-{V}olterra {M}odel with {D}emographic {N}oise},\ }\href {https://doi.org/10.1103/PhysRevLett.126.258301} {\bibfield  {journal} {\bibinfo  {journal} {Phys. Rev. Lett.}\ }\textbf {\bibinfo {volume} {126}},\ \bibinfo {pages} {258301} (\bibinfo {year} {2021})}\BibitemShut {NoStop}%
\bibitem [{\citenamefont {Al-Hiyasat}\ \emph {et~al.}(2026)\citenamefont {Al-Hiyasat}, \citenamefont {Swartz}, \citenamefont {Gore},\ and\ \citenamefont {Kardar}}]{al2026spatiotemporal}%
  \BibitemOpen
  \bibfield  {author} {\bibinfo {author} {\bibfnamefont {A.}~\bibnamefont {Al-Hiyasat}}, \bibinfo {author} {\bibfnamefont {D.~W.}\ \bibnamefont {Swartz}}, \bibinfo {author} {\bibfnamefont {J.}~\bibnamefont {Gore}},\ and\ \bibinfo {author} {\bibfnamefont {M.}~\bibnamefont {Kardar}},\ }\href@noop {} {\bibinfo {title} {Spatiotemporal noise stabilizes unbounded diversity in strongly-competitive communities}} (\bibinfo {year} {2026}),\ \Eprint {https://arxiv.org/abs/2602.13423} {arXiv:2602.13423 [q-bio.PE]} \BibitemShut {NoStop}%
\bibitem [{\citenamefont {Garcia~Lorenzana}\ \emph {et~al.}(2024)\citenamefont {Garcia~Lorenzana}, \citenamefont {Altieri},\ and\ \citenamefont {Biroli}}]{garcia2024interactions}%
  \BibitemOpen
  \bibfield  {author} {\bibinfo {author} {\bibfnamefont {G.}~\bibnamefont {Garcia~Lorenzana}}, \bibinfo {author} {\bibfnamefont {A.}~\bibnamefont {Altieri}},\ and\ \bibinfo {author} {\bibfnamefont {G.}~\bibnamefont {Biroli}},\ }\bibfield  {title} {\bibinfo {title} {Interactions and migration rescuing ecological diversity},\ }\href {https://doi.org/10.1103/PRXLife.2.013014} {\bibfield  {journal} {\bibinfo  {journal} {PRX Life}\ }\textbf {\bibinfo {volume} {2}},\ \bibinfo {pages} {013014} (\bibinfo {year} {2024})}\BibitemShut {NoStop}%
\bibitem [{\citenamefont {de~Pirey}(2025)}]{de2025self}%
  \BibitemOpen
  \bibfield  {author} {\bibinfo {author} {\bibfnamefont {T.~A.}\ \bibnamefont {de~Pirey}},\ }\href@noop {} {\bibinfo {title} {Self-organized criticality in complex model ecosystems}} (\bibinfo {year} {2025}),\ \Eprint {https://arxiv.org/abs/2512.06961} {arXiv:2512.06961 [cond-mat.stat-mech]} \BibitemShut {NoStop}%
\bibitem [{\citenamefont {Larroya}\ and\ \citenamefont {Galla}(2023)}]{larroya2023demographic}%
  \BibitemOpen
  \bibfield  {author} {\bibinfo {author} {\bibfnamefont {F.}~\bibnamefont {Larroya}}\ and\ \bibinfo {author} {\bibfnamefont {T.}~\bibnamefont {Galla}},\ }\bibfield  {title} {\bibinfo {title} {Demographic noise in complex ecological communities},\ }\href {https://doi.org/10.1088/2632-072X/acd21b} {\bibfield  {journal} {\bibinfo  {journal} {J. Phys. Complex.}\ }\textbf {\bibinfo {volume} {4}},\ \bibinfo {pages} {025012} (\bibinfo {year} {2023})}\BibitemShut {NoStop}%
\bibitem [{\citenamefont {van Kampen}(2007)}]{vankampen2007stochastic}%
  \BibitemOpen
  \bibfield  {author} {\bibinfo {author} {\bibfnamefont {N.~G.}\ \bibnamefont {van Kampen}},\ }\href@noop {} {\emph {\bibinfo {title} {Stochastic processes in physics and chemistry}}}\ (\bibinfo  {publisher} {North-Holland},\ \bibinfo {year} {2007})\BibitemShut {NoStop}%
\bibitem [{\citenamefont {M{\'e}zard}\ \emph {et~al.}(1988)\citenamefont {M{\'e}zard}, \citenamefont {Parisi}, \citenamefont {Virasoro},\ and\ \citenamefont {Thouless}}]{mezard1988spin}%
  \BibitemOpen
  \bibfield  {author} {\bibinfo {author} {\bibfnamefont {M.}~\bibnamefont {M{\'e}zard}}, \bibinfo {author} {\bibfnamefont {G.}~\bibnamefont {Parisi}}, \bibinfo {author} {\bibfnamefont {M.~A.}\ \bibnamefont {Virasoro}},\ and\ \bibinfo {author} {\bibfnamefont {D.~J.}\ \bibnamefont {Thouless}},\ }\href {https://doi.org/10.1063/1.2811676} {\bibinfo {title} {Spin glass theory and beyond}} (\bibinfo {year} {1988})\BibitemShut {NoStop}%
\bibitem [{\citenamefont {Roy}\ \emph {et~al.}(2019)\citenamefont {Roy}, \citenamefont {Biroli}, \citenamefont {Bunin},\ and\ \citenamefont {Cammarota}}]{roy2019numerical}%
  \BibitemOpen
  \bibfield  {author} {\bibinfo {author} {\bibfnamefont {F.}~\bibnamefont {Roy}}, \bibinfo {author} {\bibfnamefont {G.}~\bibnamefont {Biroli}}, \bibinfo {author} {\bibfnamefont {G.}~\bibnamefont {Bunin}},\ and\ \bibinfo {author} {\bibfnamefont {C.}~\bibnamefont {Cammarota}},\ }\bibfield  {title} {\bibinfo {title} {Numerical implementation of dynamical mean field theory for disordered systems: {A}pplication to the {L}otka--{V}olterra model of ecosystems},\ }\href {https://doi.org/10.1088/1751-8121/ab1f32} {\bibfield  {journal} {\bibinfo  {journal} {J. Phys. A: Math. and Theor.}\ }\textbf {\bibinfo {volume} {52}},\ \bibinfo {pages} {484001} (\bibinfo {year} {2019})}\BibitemShut {NoStop}%
\bibitem [{\citenamefont {Altieri}\ \emph {et~al.}(2020)\citenamefont {Altieri}, \citenamefont {Biroli},\ and\ \citenamefont {Cammarota}}]{altieri2020dynamical}%
  \BibitemOpen
  \bibfield  {author} {\bibinfo {author} {\bibfnamefont {A.}~\bibnamefont {Altieri}}, \bibinfo {author} {\bibfnamefont {G.}~\bibnamefont {Biroli}},\ and\ \bibinfo {author} {\bibfnamefont {C.}~\bibnamefont {Cammarota}},\ }\bibfield  {title} {\bibinfo {title} {Dynamical mean-field theory and aging dynamics},\ }\href {https://doi.org/10.1088/1751-8121/aba3dd} {\bibfield  {journal} {\bibinfo  {journal} {J. Phys. A: Math. Theor.}\ }\textbf {\bibinfo {volume} {53}},\ \bibinfo {pages} {375006} (\bibinfo {year} {2020})}\BibitemShut {NoStop}%
\bibitem [{\citenamefont {Ros}\ \emph {et~al.}(2023)\citenamefont {Ros}, \citenamefont {Roy}, \citenamefont {Biroli}, \citenamefont {Bunin},\ and\ \citenamefont {Turner}}]{ros2023generalized}%
  \BibitemOpen
  \bibfield  {author} {\bibinfo {author} {\bibfnamefont {V.}~\bibnamefont {Ros}}, \bibinfo {author} {\bibfnamefont {F.}~\bibnamefont {Roy}}, \bibinfo {author} {\bibfnamefont {G.}~\bibnamefont {Biroli}}, \bibinfo {author} {\bibfnamefont {G.}~\bibnamefont {Bunin}},\ and\ \bibinfo {author} {\bibfnamefont {A.~M.}\ \bibnamefont {Turner}},\ }\bibfield  {title} {\bibinfo {title} {{G}eneralized {L}otka-{V}olterra {E}quations with {R}andom, {N}onreciprocal {I}nteractions: {T}he {T}ypical {N}umber of {E}quilibria},\ }\href {https://doi.org/10.1103/PhysRevLett.130.257401} {\bibfield  {journal} {\bibinfo  {journal} {Phys. Rev. Lett.}\ }\textbf {\bibinfo {volume} {130}},\ \bibinfo {pages} {257401} (\bibinfo {year} {2023})}\BibitemShut {NoStop}%
\bibitem [{\citenamefont {Krumbeck}\ \emph {et~al.}(2021)\citenamefont {Krumbeck}, \citenamefont {Yang}, \citenamefont {Constable},\ and\ \citenamefont {Rogers}}]{krumbeck2021fluctuation}%
  \BibitemOpen
  \bibfield  {author} {\bibinfo {author} {\bibfnamefont {Y.}~\bibnamefont {Krumbeck}}, \bibinfo {author} {\bibfnamefont {Q.}~\bibnamefont {Yang}}, \bibinfo {author} {\bibfnamefont {G.~W.~A.}\ \bibnamefont {Constable}},\ and\ \bibinfo {author} {\bibfnamefont {T.}~\bibnamefont {Rogers}},\ }\bibfield  {title} {\bibinfo {title} {Fluctuation spectra of large random dynamical systems reveal hidden structure in ecological networks},\ }\href {https://doi.org/10.1038/s41467-021-23757-x} {\bibfield  {journal} {\bibinfo  {journal} {Nat. Commun.}\ }\textbf {\bibinfo {volume} {12}},\ \bibinfo {pages} {3625} (\bibinfo {year} {2021})}\BibitemShut {NoStop}%
\bibitem [{\citenamefont {Allesina}\ and\ \citenamefont {Tang}(2012)}]{allesina2012stability}%
  \BibitemOpen
  \bibfield  {author} {\bibinfo {author} {\bibfnamefont {S.}~\bibnamefont {Allesina}}\ and\ \bibinfo {author} {\bibfnamefont {S.}~\bibnamefont {Tang}},\ }\bibfield  {title} {\bibinfo {title} {Stability criteria for complex ecosystems},\ }\href {https://doi.org/10.1038/nature10832} {\bibfield  {journal} {\bibinfo  {journal} {Nature}\ }\textbf {\bibinfo {volume} {483}},\ \bibinfo {pages} {205} (\bibinfo {year} {2012})}\BibitemShut {NoStop}%
\bibitem [{\citenamefont {Baron}\ \emph {et~al.}(2023)\citenamefont {Baron}, \citenamefont {Jewell}, \citenamefont {Ryder},\ and\ \citenamefont {Galla}}]{baron2023breakdown}%
  \BibitemOpen
  \bibfield  {author} {\bibinfo {author} {\bibfnamefont {J.~W.}\ \bibnamefont {Baron}}, \bibinfo {author} {\bibfnamefont {T.~J.}\ \bibnamefont {Jewell}}, \bibinfo {author} {\bibfnamefont {C.}~\bibnamefont {Ryder}},\ and\ \bibinfo {author} {\bibfnamefont {T.}~\bibnamefont {Galla}},\ }\bibfield  {title} {\bibinfo {title} {Breakdown of random-matrix universality in persistent {L}otka-{V}olterra communities},\ }\href {https://doi.org/10.1103/PhysRevLett.130.137401} {\bibfield  {journal} {\bibinfo  {journal} {Phys. Rev. Lett.}\ }\textbf {\bibinfo {volume} {130}},\ \bibinfo {pages} {137401} (\bibinfo {year} {2023})}\BibitemShut {NoStop}%
\bibitem [{\citenamefont {Biroli}\ \emph {et~al.}(2018)\citenamefont {Biroli}, \citenamefont {Bunin},\ and\ \citenamefont {Cammarota}}]{biroli2018marginally}%
  \BibitemOpen
  \bibfield  {author} {\bibinfo {author} {\bibfnamefont {G.}~\bibnamefont {Biroli}}, \bibinfo {author} {\bibfnamefont {G.}~\bibnamefont {Bunin}},\ and\ \bibinfo {author} {\bibfnamefont {C.}~\bibnamefont {Cammarota}},\ }\bibfield  {title} {\bibinfo {title} {Marginally stable equilibria in critical ecosystems},\ }\href {https://doi.org/10.1088/1367-2630/aada58} {\bibfield  {journal} {\bibinfo  {journal} {New J. Phys.}\ }\textbf {\bibinfo {volume} {20}},\ \bibinfo {pages} {083051} (\bibinfo {year} {2018})}\BibitemShut {NoStop}%
\bibitem [{\citenamefont {White}\ \emph {et~al.}(2012)\citenamefont {White}, \citenamefont {Thibault},\ and\ \citenamefont {Xiao}}]{white2012characterizing}%
  \BibitemOpen
  \bibfield  {author} {\bibinfo {author} {\bibfnamefont {E.~P.}\ \bibnamefont {White}}, \bibinfo {author} {\bibfnamefont {K.~M.}\ \bibnamefont {Thibault}},\ and\ \bibinfo {author} {\bibfnamefont {X.}~\bibnamefont {Xiao}},\ }\bibfield  {title} {\bibinfo {title} {Characterizing species abundance distributions across taxa and ecosystems using a simple maximum entropy model},\ }\href {https://doi.org/10.1890/11-2177.1} {\bibfield  {journal} {\bibinfo  {journal} {Ecology}\ }\textbf {\bibinfo {volume} {93}},\ \bibinfo {pages} {1772} (\bibinfo {year} {2012})}\BibitemShut {NoStop}%
\bibitem [{\citenamefont {Locey}\ and\ \citenamefont {Lennon}(2016)}]{locey2016scaling}%
  \BibitemOpen
  \bibfield  {author} {\bibinfo {author} {\bibfnamefont {K.~J.}\ \bibnamefont {Locey}}\ and\ \bibinfo {author} {\bibfnamefont {J.~T.}\ \bibnamefont {Lennon}},\ }\bibfield  {title} {\bibinfo {title} {Scaling laws predict global microbial diversity},\ }\href {https://doi.org/10.1073/pnas.1521291113} {\bibfield  {journal} {\bibinfo  {journal} {Proc. Natl. Acad. Sci. U.S.A.}\ }\textbf {\bibinfo {volume} {113}},\ \bibinfo {pages} {5970} (\bibinfo {year} {2016})}\BibitemShut {NoStop}%
\bibitem [{\citenamefont {H{\"a}nggi}\ \emph {et~al.}(1990)\citenamefont {H{\"a}nggi}, \citenamefont {Talkner},\ and\ \citenamefont {Borkovec}}]{hanggi1990reaction}%
  \BibitemOpen
  \bibfield  {author} {\bibinfo {author} {\bibfnamefont {P.}~\bibnamefont {H{\"a}nggi}}, \bibinfo {author} {\bibfnamefont {P.}~\bibnamefont {Talkner}},\ and\ \bibinfo {author} {\bibfnamefont {M.}~\bibnamefont {Borkovec}},\ }\bibfield  {title} {\bibinfo {title} {Reaction-rate theory: fifty years after kramers},\ }\href@noop {} {\bibfield  {journal} {\bibinfo  {journal} {Reviews of modern physics}\ }\textbf {\bibinfo {volume} {62}},\ \bibinfo {pages} {251} (\bibinfo {year} {1990})}\BibitemShut {NoStop}%
\bibitem [{\citenamefont {Assaf}\ and\ \citenamefont {Meerson}(2017)}]{assaf2017wkb}%
  \BibitemOpen
  \bibfield  {author} {\bibinfo {author} {\bibfnamefont {M.}~\bibnamefont {Assaf}}\ and\ \bibinfo {author} {\bibfnamefont {B.}~\bibnamefont {Meerson}},\ }\bibfield  {title} {\bibinfo {title} {{WKB} theory of large deviations in stochastic populations},\ }\href {https://doi.org/10.1088/1751-8121/aa669a} {\bibfield  {journal} {\bibinfo  {journal} {J. Phys. A: Math. Theor.}\ }\textbf {\bibinfo {volume} {50}},\ \bibinfo {pages} {263001} (\bibinfo {year} {2017})}\BibitemShut {NoStop}%
\bibitem [{\citenamefont {Bressloff}(2022)}]{bressloff2022wkb}%
  \BibitemOpen
  \bibfield  {author} {\bibinfo {author} {\bibfnamefont {P.~C.}\ \bibnamefont {Bressloff}},\ }\href@noop {} {\emph {\bibinfo {title} {Stochastic Processes in Cell Biology: Volume I}}}\ (\bibinfo  {publisher} {Springer},\ \bibinfo {year} {2022})\BibitemShut {NoStop}%
\bibitem [{\citenamefont {Ros}\ \emph {et~al.}(2021)\citenamefont {Ros}, \citenamefont {Biroli},\ and\ \citenamefont {Cammarota}}]{ros2021dynamical}%
  \BibitemOpen
  \bibfield  {author} {\bibinfo {author} {\bibfnamefont {V.}~\bibnamefont {Ros}}, \bibinfo {author} {\bibfnamefont {G.}~\bibnamefont {Biroli}},\ and\ \bibinfo {author} {\bibfnamefont {C.}~\bibnamefont {Cammarota}},\ }\bibfield  {title} {\bibinfo {title} {Dynamical instantons and activated processes in mean-field glass models},\ }\href {https://doi.org/10.21468/SciPostPhys.10.1.002} {\bibfield  {journal} {\bibinfo  {journal} {SciPost Phys.}\ }\textbf {\bibinfo {volume} {10}},\ \bibinfo {pages} {002} (\bibinfo {year} {2021})}\BibitemShut {NoStop}%
\bibitem [{\citenamefont {Lopatin}\ and\ \citenamefont {Ioffe}(1999)}]{lopatin1999instantons}%
  \BibitemOpen
  \bibfield  {author} {\bibinfo {author} {\bibfnamefont {A.~V.}\ \bibnamefont {Lopatin}}\ and\ \bibinfo {author} {\bibfnamefont {L.~B.}\ \bibnamefont {Ioffe}},\ }\bibfield  {title} {\bibinfo {title} {Instantons in the {L}angevin dynamics: {A}n application to spin glasses},\ }\href {https://doi.org/10.1103/PhysRevB.60.6412} {\bibfield  {journal} {\bibinfo  {journal} {Phys. Rev. B}\ }\textbf {\bibinfo {volume} {60}},\ \bibinfo {pages} {6412} (\bibinfo {year} {1999})}\BibitemShut {NoStop}%
\bibitem [{\citenamefont {Poley}\ \emph {et~al.}(2023)\citenamefont {Poley}, \citenamefont {Baron},\ and\ \citenamefont {Galla}}]{poley2023generalized}%
  \BibitemOpen
  \bibfield  {author} {\bibinfo {author} {\bibfnamefont {L.}~\bibnamefont {Poley}}, \bibinfo {author} {\bibfnamefont {J.~W.}\ \bibnamefont {Baron}},\ and\ \bibinfo {author} {\bibfnamefont {T.}~\bibnamefont {Galla}},\ }\bibfield  {title} {\bibinfo {title} {Generalized {L}otka-{V}olterra model with hierarchical interactions},\ }\href {https://doi.org/10.1103/PhysRevE.107.024313} {\bibfield  {journal} {\bibinfo  {journal} {Phys. Rev. E}\ }\textbf {\bibinfo {volume} {107}},\ \bibinfo {pages} {024313} (\bibinfo {year} {2023})}\BibitemShut {NoStop}%
\bibitem [{\citenamefont {Grilli}\ \emph {et~al.}(2016)\citenamefont {Grilli}, \citenamefont {Rogers},\ and\ \citenamefont {Allesina}}]{grilli2016modularity}%
  \BibitemOpen
  \bibfield  {author} {\bibinfo {author} {\bibfnamefont {J.}~\bibnamefont {Grilli}}, \bibinfo {author} {\bibfnamefont {T.}~\bibnamefont {Rogers}},\ and\ \bibinfo {author} {\bibfnamefont {S.}~\bibnamefont {Allesina}},\ }\bibfield  {title} {\bibinfo {title} {Modularity and stability in ecological communities},\ }\href {https://doi.org/10.1038/ncomms12031} {\bibfield  {journal} {\bibinfo  {journal} {Nat. Commun.}\ }\textbf {\bibinfo {volume} {7}},\ \bibinfo {pages} {12031} (\bibinfo {year} {2016})}\BibitemShut {NoStop}%
\bibitem [{\citenamefont {Allesina}\ \emph {et~al.}(2015)\citenamefont {Allesina}, \citenamefont {Grilli}, \citenamefont {Barab{\'a}s}, \citenamefont {Tang}, \citenamefont {Aljadeff},\ and\ \citenamefont {Maritan}}]{allesina2015predicting}%
  \BibitemOpen
  \bibfield  {author} {\bibinfo {author} {\bibfnamefont {S.}~\bibnamefont {Allesina}}, \bibinfo {author} {\bibfnamefont {J.}~\bibnamefont {Grilli}}, \bibinfo {author} {\bibfnamefont {G.}~\bibnamefont {Barab{\'a}s}}, \bibinfo {author} {\bibfnamefont {S.}~\bibnamefont {Tang}}, \bibinfo {author} {\bibfnamefont {J.}~\bibnamefont {Aljadeff}},\ and\ \bibinfo {author} {\bibfnamefont {A.}~\bibnamefont {Maritan}},\ }\bibfield  {title} {\bibinfo {title} {Predicting the stability of large structured food webs},\ }\href {https://doi.org/10.1038/ncomms8842} {\bibfield  {journal} {\bibinfo  {journal} {Nat. Commun.}\ }\textbf {\bibinfo {volume} {6}},\ \bibinfo {pages} {7842} (\bibinfo {year} {2015})}\BibitemShut {NoStop}%
\bibitem [{\citenamefont {Barab{\'a}s}\ \emph {et~al.}(2017)\citenamefont {Barab{\'a}s}, \citenamefont {Michalska-Smith},\ and\ \citenamefont {Allesina}}]{barabas2017self}%
  \BibitemOpen
  \bibfield  {author} {\bibinfo {author} {\bibfnamefont {G.}~\bibnamefont {Barab{\'a}s}}, \bibinfo {author} {\bibfnamefont {M.~J.}\ \bibnamefont {Michalska-Smith}},\ and\ \bibinfo {author} {\bibfnamefont {S.}~\bibnamefont {Allesina}},\ }\bibfield  {title} {\bibinfo {title} {Self-regulation and the stability of large ecological networks},\ }\href {https://doi.org/10.1038/s41559-017-0357-6} {\bibfield  {journal} {\bibinfo  {journal} {Nat. Ecol. Evol.}\ }\textbf {\bibinfo {volume} {1}},\ \bibinfo {pages} {1870} (\bibinfo {year} {2017})}\BibitemShut {NoStop}%
\bibitem [{\citenamefont {Gardiner}(2009)}]{gardiner2009stochastic}%
  \BibitemOpen
  \bibfield  {author} {\bibinfo {author} {\bibfnamefont {C.}~\bibnamefont {Gardiner}},\ }\href@noop {} {\emph {\bibinfo {title} {Stochastic Methods}}}\ (\bibinfo  {publisher} {Springer},\ \bibinfo {year} {2009})\BibitemShut {NoStop}%
\bibitem [{\citenamefont {Kurtz}(1978)}]{kurtz1978strong}%
  \BibitemOpen
  \bibfield  {author} {\bibinfo {author} {\bibfnamefont {T.~G.}\ \bibnamefont {Kurtz}},\ }\bibfield  {title} {\bibinfo {title} {{S}trong approximation theorems for density dependent {M}arkov chains},\ }\href {https://doi.org/https://doi.org/10.1016/0304-4149(78)90020-0} {\bibfield  {journal} {\bibinfo  {journal} {Stoc. Process. Appl.}\ }\textbf {\bibinfo {volume} {6}},\ \bibinfo {pages} {223} (\bibinfo {year} {1978})}\BibitemShut {NoStop}%
\end{thebibliography}

\begin{thebibliography}{7}%
\makeatletter
\providecommand \@ifxundefined [1]{%
 \@ifx{#1\undefined}
}%
\providecommand \@ifnum [1]{%
 \ifnum #1\expandafter \@firstoftwo
 \else \expandafter \@secondoftwo
 \fi
}%
\providecommand \@ifx [1]{%
 \ifx #1\expandafter \@firstoftwo
 \else \expandafter \@secondoftwo
 \fi
}%
\providecommand \natexlab [1]{#1}%
\providecommand \enquote  [1]{``#1''}%
\providecommand \bibnamefont  [1]{#1}%
\providecommand \bibfnamefont [1]{#1}%
\providecommand \citenamefont [1]{#1}%
\providecommand \href@noop [0]{\@secondoftwo}%
\providecommand \href [0]{\begingroup \@sanitize@url \@href}%
\providecommand \@href[1]{\@@startlink{#1}\@@href}%
\providecommand \@@href[1]{\endgroup#1\@@endlink}%
\providecommand \@sanitize@url [0]{\catcode `\\12\catcode `\$12\catcode `\&12\catcode `\#12\catcode `\^12\catcode `\_12\catcode `\%12\relax}%
\providecommand \@@startlink[1]{}%
\providecommand \@@endlink[0]{}%
\providecommand \url  [0]{\begingroup\@sanitize@url \@url }%
\providecommand \@url [1]{\endgroup\@href {#1}{\urlprefix }}%
\providecommand \urlprefix  [0]{URL }%
\providecommand \Eprint [0]{\href }%
\providecommand \doibase [0]{https://doi.org/}%
\providecommand \selectlanguage [0]{\@gobble}%
\providecommand \bibinfo  [0]{\@secondoftwo}%
\providecommand \bibfield  [0]{\@secondoftwo}%
\providecommand \translation [1]{[#1]}%
\providecommand \BibitemOpen [0]{}%
\providecommand \bibitemStop [0]{}%
\providecommand \bibitemNoStop [0]{.\EOS\space}%
\providecommand \EOS [0]{\spacefactor3000\relax}%
\providecommand \BibitemShut  [1]{\csname bibitem#1\endcsname}%
\let\auto@bib@innerbib\@empty
\bibitem [{\citenamefont {van Kampen}(2007)}]{Svankampen2007stochastic}%
  \BibitemOpen
  \bibfield  {author} {\bibinfo {author} {\bibfnamefont {N.~G.}\ \bibnamefont {van Kampen}},\ }\href@noop {} {\emph {\bibinfo {title} {Stochastic processes in physics and chemistry}}}\ (\bibinfo  {publisher} {North-Holland},\ \bibinfo {year} {2007})\BibitemShut {NoStop}%
\bibitem [{\citenamefont {Gardiner}(2009)}]{Sgardiner2009stochastic}%
  \BibitemOpen
  \bibfield  {author} {\bibinfo {author} {\bibfnamefont {C.}~\bibnamefont {Gardiner}},\ }\href@noop {} {\emph {\bibinfo {title} {Stochastic Methods}}}\ (\bibinfo  {publisher} {Springer},\ \bibinfo {year} {2009})\BibitemShut {NoStop}%
\bibitem [{\citenamefont {Kurtz}(1978)}]{Skurtz1978strong}%
  \BibitemOpen
  \bibfield  {author} {\bibinfo {author} {\bibfnamefont {T.~G.}\ \bibnamefont {Kurtz}},\ }\bibfield  {title} {\bibinfo {title} {{S}trong approximation theorems for density dependent {M}arkov chains},\ }\href {https://doi.org/https://doi.org/10.1016/0304-4149(78)90020-0} {\bibfield  {journal} {\bibinfo  {journal} {Stoc. Process. Appl.}\ }\textbf {\bibinfo {volume} {6}},\ \bibinfo {pages} {223} (\bibinfo {year} {1978})}\BibitemShut {NoStop}%
\bibitem [{\citenamefont {Gillespie}(2000)}]{Sgillespie2000chemical}%
  \BibitemOpen
  \bibfield  {author} {\bibinfo {author} {\bibfnamefont {D.~T.}\ \bibnamefont {Gillespie}},\ }\bibfield  {title} {\bibinfo {title} {The chemical {L}angevin equation},\ }\href {https://doi.org/10.1063/1.481811} {\bibfield  {journal} {\bibinfo  {journal} {J. Chem. Phys.}\ }\textbf {\bibinfo {volume} {113}},\ \bibinfo {pages} {297} (\bibinfo {year} {2000})}\BibitemShut {NoStop}%
\bibitem [{\citenamefont {Galla}(2024)}]{Sgalla2024generating}%
  \BibitemOpen
  \bibfield  {author} {\bibinfo {author} {\bibfnamefont {T.}~\bibnamefont {Galla}},\ }\href {https://arxiv.org/abs/2405.14289} {\bibinfo {title} {Generating-functional analysis of random {L}otka-{V}olterra systems: {A} step-by-step guide}} (\bibinfo {year} {2024}),\ \Eprint {https://arxiv.org/abs/2405.14289} {arXiv:2405.14289 [cond-mat.dis-nn]} \BibitemShut {NoStop}%
\bibitem [{\citenamefont {Baron}(2026)}]{Sbaron2026lecture}%
  \BibitemOpen
  \bibfield  {author} {\bibinfo {author} {\bibfnamefont {J.~W.}\ \bibnamefont {Baron}},\ }\href {https://arxiv.org/abs/2607.07868} {\bibinfo {title} {Lecture notes on random matrix theory: the results, the applications, and the analytical tools}} (\bibinfo {year} {2026}),\ \Eprint {https://arxiv.org/abs/2607.07868} {arXiv:2607.07868 [cond-mat.dis-nn]} \BibitemShut {NoStop}%
\bibitem [{\citenamefont {Assaf}\ and\ \citenamefont {Meerson}(2017)}]{Sassaf2017wkb}%
  \BibitemOpen
  \bibfield  {author} {\bibinfo {author} {\bibfnamefont {M.}~\bibnamefont {Assaf}}\ and\ \bibinfo {author} {\bibfnamefont {B.}~\bibnamefont {Meerson}},\ }\bibfield  {title} {\bibinfo {title} {{WKB} theory of large deviations in stochastic populations},\ }\href {https://doi.org/10.1088/1751-8121/aa669a} {\bibfield  {journal} {\bibinfo  {journal} {J. Phys. A: Math. Theor.}\ }\textbf {\bibinfo {volume} {50}},\ \bibinfo {pages} {263001} (\bibinfo {year} {2017})}\BibitemShut {NoStop}%
\end{thebibliography}
\end{document}